\documentclass[a4paper,11pt]{article}
\usepackage{jcappub} 
\usepackage{lineno}
\usepackage{subcaption}
\usepackage{graphicx}
\usepackage{amsmath}
\usepackage{booktabs}
\usepackage{siunitx}
\usepackage{mathtools}
\usepackage[normalem]{ulem}
\usepackage[shortlabels]{enumitem}
\usepackage[symbol]{footmisc}
\DeclareSIUnit\degsymbol{\degree}
\DeclareSIUnit\Msun{M_{\odot}}
\DeclareSIUnit\yr{yr}
\DeclareSIUnit\pc{pc}
\DeclareSIUnit\kpc{kpc}
\DeclareSIUnit\Mpc{Mpc}

\usepackage{xcolor}
\def\code#1{\texttt{#1}}

\title{Gamma rays from dark matter spikes in EAGLE simulations}

\author[a,b,*]{J. Aschersleben,\note{Corresponding author.}}
\author[c]{G. Bertone,}
\author[d]{D. Horns,}
\author[e]{E. Moulin,}
\author[a]{R. F. Peletier}
\author[a]{and M. Vecchi}

\affiliation[a]{Kapteyn Astronomical Institute, University of Groningen, PO Box 800, NL-9700 AV, Groningen, The Netherlands}
\affiliation[b]{Bernoulli Institute for Mathematics, Computer Science and Artificial Intelligence, University of Groningen, PO Box 407, NL-9700 AK, Groningen, The Netherlands}
\affiliation[c]{Gravitation Astroparticle Physics Amsterdam (GRAPPA), University of Amsterdam, Science Park 904, 1098 XH Amsterdam, The Netherlands}
\affiliation[d]{Institut für Experimentalphysik, University of Hamburg, Luruper Chaussee 149, 22761 Hamburg, Germany}
\affiliation[e]{IRFU, CEA, Université Paris-Saclay, F-91191 Gif-sur-Yvette, France}

\emailAdd{j.j.m.aschersleben@rug.nl}
\emailAdd{g.bertone@uva.nl}
\emailAdd{dieter.horns@uni-hamburg.de}
\emailAdd{emmanuel.moulin@cea.fr}
\emailAdd{r.f.peletier@rug.nl}
\emailAdd{m.vecchi@rug.nl}

\abstract{Intermediate Mass Black Holes (IMBHs) with a mass range between $\SI{100}{\Msun}$ and $\SI{e6}{\Msun}$ are expected to be surrounded by high dark matter densities, so-called dark matter spikes. 
The high density of self-annihilating Weakly Interacting Massive Particles (WIMPs) in these spikes leads to copious gamma-ray production. Sufficiently nearby IMBHs could therefore appear as unidentified gamma-ray sources.
However, the number of IMBHs and their distribution within our own Milky Way is currently unknown. In this work, we provide a mock catalogue of IMBHs and their dark matter spikes obtained from the EAGLE simulations, in which black holes with a mass of $\SI{e5}{\Msun} /h$ are seeded into the centre of halos greater than $\SI{e10}{\Msun} /h$ to model black hole feedback influencing the formation of galaxies.
The catalogue contains the coordinates and dark matter spike parameters for about 2500 IMBHs present in about 150 Milky Way-like galaxies. We expect about $15^{+9}_{-6}$ IMBHs within our own galaxy, mainly distributed in the Galactic Centre and the Galactic Plane. In the most optimistic scenario, we find that current and future gamma-ray observatories, such as Fermi-LAT, H.E.S.S. and CTAO, would be sensitive enough to probe the cross section of dark matter self-annihilation around IMBHs down to many orders of magnitude below the thermal relic cross section for dark matter particles with masses from GeV to TeV.
We have made the IMBH mock catalogue and the source code for our analysis publicly available, providing the resources to study dark matter self-annihilation around IMBHs with current and upcoming gamma-ray observatories.}

\begin{document}
\maketitle
\flushbottom

\section{Introduction} \label{sec:introduction}
The nature of dark matter is one of the most important questions in fundamental physics~\cite{bertone2018new,boveia2018dark,gaskins2016review,Mayet2016zxu}. 
One of the most popular candidates for dark matter are Weakly Interacting Massive Particles (WIMPs), which naturally arise in well-motivated extensions of the Standard Model of particle physics ~\cite{bertone2005particle,bertone2010particle,arcadi2018waning,leane2018gev}.
The thermal production of WIMPs at the measured relic density of $\Omega_{\mathrm{DM}} = 0.26$ implies the production of Standard Model particles, including gamma rays, neutrinos, and anti-particles through the self-annihilation cross section $\Omega_{\mathrm{DM}} h^2 \approx 3 \times 10^{-27}\mathrm{cm^3s^{-1}} / \langle \sigma v \rangle$~\cite{ade2014planck,profumo2017introduction}. This canonical annihilation cross section, approximated at the time of chemical decoupling, is applicable primarily for indirect detection scenarios in case of velocity-independent (\textit{s}-wave) processes, and may not necessarily be probed by direct dark matter searches. The indirect detection of self-annihilating dark matter includes searches for gamma-ray emission from regions with large WIMP number densities $n_\mathrm{WIMP}$, as the self-annihilation rate scales with $n_\mathrm{WIMP}^2$.
A lot of effort has been put into searching for a dark matter signal in regions with large dark matter number densities, including the Galactic Centre~\cite{abramowski2011search,hooper2011dark,albert2006observation, Abdallah:2016ygi, abeysekara2018search,HESS:2022ygk}, dwarf galaxies~\cite{abramowski2014search,ahnen2016limits,zitzer2015search,HESS:2020zwn} and galaxy clusters~\cite{abramowski2012search,huang2012probing,aleksic2010magic}.
Another very promising target for indirect dark matter searches are the environments of black holes dominated by their gravitational potential.
Gondolo and Silk (1999)~\cite{gondolo1999dark} applied the formalism developed for the adiabatic growth of  power-law stellar cusps by Quinlan, Hernquist and Sigurdsson (1995) \cite{1995ApJ...440..554Q}, to the dark matter density profile around the supermassive black hole (SMBH) at the Galactic Centre, Sgr~A$^*$.
Since the dark matter annihilation rate is proportional to the squared dark matter density $\rho_\mathrm{DM}^2$, these adiabatically grown profiles, dubbed \textit{spikes}, would lead to a significant enhancement of the gamma-ray signal. Subsequent studies explored the implications of the existence of spikes around SMBHs~\cite{2002MNRAS.337...98B,gnedin2004dark}, deriving constraints on the dark matter cross section based on the spike around Sgr~A$^*$~\cite{2002PhRvL..88s1301M,2005PhRvD..72j3502B,2002MNRAS.337...98B,2004JCAP...05..007A,2004PhRvD..70k3007H,fields2014galactic,balaji2023dark}.
Zhao and Silk (2005)~\cite{zhao2005dark} and Bertone, Zentner, and Silk (2005)~\cite{bertone2005new}, proposed spikes around Intermediate Mass Black Holes (IMBHs)~\cite{greene2020intermediate}  as promising targets for indirect dark matter searches. 
IMBHs cover a mass range between $\sim \SI{e2}{}-\SI{e6}{\Msun}$ and are expected to form via gravitational runaway~\cite{rosswog2009tidal}, the direct collapse of primordial gas in early forming halos~\cite{loeb1994collapse}, as remnants of Population III stars~\cite{karlsson2013pregalactic,madau2001massive} or from primordial black holes (PBHs)~\cite{kawaguchi2008formation}. Potential IMBH candidates have been found in ultraluminous X-ray sources (ULXs)~\cite{long1981observations}. These sources have luminosities above the Eddington limit for compact objects with $M \lesssim \SI{20}{\Msun}$ and can therefore not be explained by neutron stars and stellar mass black holes~\cite{greene2020intermediate}. The most well-studied and recognized systems are $\omega$~Centauri and 47 Tucanae, for which black hole masses in the order of $\SI{e4}{\Msun}$ have been found~\cite{noyola2010very, kiziltan2017intermediate}. However, additional observational data are needed to confirm these measurements~\cite{greene2020intermediate}. The first conclusive evidence for an IMBH is the detection of the gravitational waves from the binary black hole merger event GW190521 with an inferred mass of the remnant black hole of $142^{+28}_{-16} \SI{}{\Msun}$~\cite{abbott2020gw190521}.
However, the observable black hole mass range of these detectors is currently limited, spanning from a few solar masses up to a couple of hundred solar masses~\cite{abbott2023population}. Therefore, current detectors do not allow to probe IMBHs in the $M \gtrsim \SI{e3}{\Msun}$ mass range. The upcoming LISA experiment~\cite{amaro2017laser} will cover the $10^{-4}-\SI{e-2}{\Hz}$ frequency range of gravitational waves enabling the detection of IMBHs above the $M \gtrsim \SI{e3}{\Msun}$ mass regime~\cite{miller2009intermediate}. \\
The distributions and luminosity of IMBHs and their dark matter spikes have been estimated by Bertone, Zentner and Silk (2005)~\cite{bertone2005new}, showing that the spikes may be detected as unidentified point-like gamma-ray sources. Dark matter spikes around IMBHs in the Milky Way would result in tens or more point-like sources with identical energy spectra, which would make them a smoking gun signature for dark matter annihilation~\cite{bertone2005new,bertone2009dark}.
Early gamma-ray searches from dark matter annihilation around IMBHs have been reported in Aharonian et al. (2008)~\cite{aharonian2008search} and Bringmann, Lavalle and Salati (2009)~\cite{bringmann2009intermediate}.
Nowadays, the advancements in cosmological simulations, such as the EAGLE~\cite{schaye2015eagle} and IllustrisTNG~\cite{pillepich2018simulating} simulations, provide a more comprehensive and refined understanding of the impact of massive black holes and their associated phenomena within their host galaxies. However, it is important to acknowledge the substantial theoretical uncertainties associated with the formation and evolution of IMBHs inherent in these simulations. Recent studies, such as Fujii et al. (2024)~\cite{fujii2024simulations}, have highlighted the complexities of IMBH formation, particularly the intricate baryonic processes that occur on small scales~\cite{giersz2015mocca,fragione2022repeated,rizzuto2021intermediate}. In their work, Fujii et al. conducted high-resolution hydrodynamic simulations of giant molecular clouds, demonstrating how IMBHs can form from the mergers of very massive stars in dense star clusters, where both the total mass and the metallicity of the cloud play crucial roles. These studies suggest that current cosmological simulations do not fully capture these detailed processes due to their limited spatial and temporal resolutions. Nevertheless, integrating insights from cosmological simulations allows us to build a more comprehensive picture of the distribution of IMBHs and their corresponding dark matter spikes within the Milky Way. \\
In this work, we provide a mock catalogue of IMBHs within Milky Way-like galaxies using the EAGLE simulations~\cite{schaye2015eagle}. The catalogue provides information about the expected number of IMBHs, their spatial distribution and their dark matter spike parameters, including the expected gamma-ray flux from dark matter self-annihilation. The IMBH catalogue, the catalogue of our selection of Milky Way-like galaxies within the EAGLE simulations and the source code are publicly available at~\cite{aschersleben_2024_10566372} and~\cite{aschersleben_2024_10491705}. We perform our analysis within the framework of a flat $\Lambda$ Cold Dark Matter ($\Lambda$CDM) cosmology, following the parameters from the Planck mission~\cite{ade2014planck}, i.e. $\Omega_{\mathrm{m}} = 0.307$, $\Omega_{\Lambda} = 0.693$, $\Omega_{\mathrm{b}} = 0.04825$, $h = 0.6777$, $\sigma_8 = 0.8288$ and $n_{\mathrm{s}} = 0.9611$. \\
The structure of this article is as follows: In Section~\ref{sec:spikes}, we introduce the theoretical background of the dark matter spikes and the gamma-ray flux expected from dark matter annihilation. We describe the EAGLE simulations and our selection criteria for Milky Way-like galaxies in Section~\ref{sec:dataset}. The properties of selected EAGLE galaxies, the analysis steps to extract the IMBH coordinates and the dark matter spike parameters for each Milky Way-like galaxy are described in Section~\ref{sec:analysis}. We present our results and discussion for the detectability of a dark matter annihilation signal around IMBHs in Section~\ref{sec:results} \& Section~\ref{sec:discussion}. Finally, we conclude our findings in Section~\ref{sec:conclusion}. \\

\section{Dark matter spikes}\label{sec:spikes}
We start with describing the theoretical framework of dark matter spikes around intermediate mass black holes. This framework is used to calculate the expected gamma-ray flux from dark matter self-annihilation around IMBHs in Section~\ref{sec:analysis} \&~\ref{sec:results}.

\subsection{Dark matter density surrounding IMBHs}
\begin{figure}
    \centering
    \includegraphics{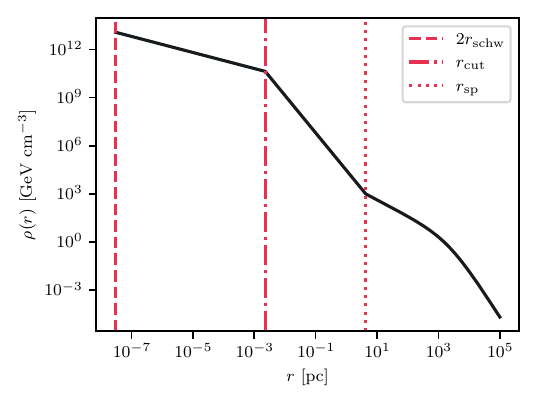}
    \caption{Dark matter density profile around an IMBH assuming a black hole mass of $\SI{1.6e5}{\Msun}$, the NFW profile as dark matter halo profile with $\rho_0 = \SI{1.9}{\GeV \per \cubic \cm}$ and $r_\mathrm{s} = \SI{2.3}{\kpc}$, a spike radius of $r_\mathrm{sp} = \SI{4.2}{\pc}$ and cutoff radius of $r_\mathrm{cut} = \SI{2.3e-3}{\pc}$. These values correspond to the parameters from a typical IMBH in our analysis.}
    \label{fig:spike_profile}
\end{figure}

We follow the theoretical framework of Bertone, Zentner and Silk (2005)~\cite{bertone2005new} to calculate the dark matter density profile surrounding IMBHs. We parametrise the dark matter density profile as follows
\begin{equation}
    \label{eq:spikedensity}
        \rho(r)= \left\{
            \begin{array}{ll}
                0 & \quad r\leq 2 r_\mathrm{schw} \\
                \rho_\mathrm{wcusp}(r) & \quad 2 r_\mathrm{schw} < r \leq r_\textrm{cut} \\
                \displaystyle\rho_\textrm{sp}(r) & \quad r_\textrm{cut} < r \leq r_\textrm{sp} \\
                \rho_\textrm{halo}(r) & \quad r > r_\textrm{sp}
            \end{array}
        \right.
\end{equation}
with the Schwarzschild radius $r_\mathrm{schw} = 2Gm_\mathrm{BH}/c^2$, the mass of the black hole $m_\mathrm{BH}$, the spike radius $r_\mathrm{sp}$, the cutoff radius $r_\mathrm{cut}$, the dark matter weak cusp density profile $\rho_\mathrm{wcusp}(r)$, the dark matter spike density profile $\rho_\mathrm{sp}(r)$ and the dark matter density profile of the host halo $\rho_\mathrm{halo}(r)$, in which the black hole formed. Therefore, the dark matter density profile around an IMBH consists of four regions: 1.) the region inside $2r_\mathrm{schw}$, in which we assume the dark matter density to be zero, 2.) a weak cusp between $2r_\mathrm{schw}$ and the cutoff radius, characterised by $r^{-0.5}$ due to dark matter \textit{s}-wave annihilation~\cite{vasiliev2007dark,shapiro2016weak},
3.) the region between the cutoff radius and the spike radius, which corresponds to the dark matter spike and 4.) the region outside the spike radius, which follows the dark matter density profile of the host halo. We discuss these individual components and how to compute the dark matter spike parameters in the following. \\
In N-body cosmological simulations, the Navarro-Frenk-White (NFW) profile has been shown to describe the dark matter density profile of galaxies very well~\cite{navarro1997universal}. Therefore, we assume that the dark matter density profile of the host halo $\rho_\mathrm{halo}(r)$ follows a NFW profile~\cite{navarro1997universal} given by
\begin{equation}\label{eq:nfw}
    \rho_\mathrm{halo}(r) = \rho_\mathrm{NFW}(r) = \rho_0 \left(\frac{r}{r_\mathrm{s}}\right)^{-1} \left(1+\frac{r}{r_\mathrm{s}}\right)^{-2} 
\end{equation}
with the normalisation constant $\rho_0$ and the scale radius $r_\mathrm{s}$.
The dark matter density profile $\rho_\mathrm{NFW}(r)$ is used to calculate the radius of the sphere of gravitational influence $r_\mathrm{h}$ of the black hole which is given by~\cite{merritt2003single}
\begin{equation}\label{eq:r_h}
    M(<r_h) = 4 \pi \int^{r_\mathrm{h}}_0 \rho_\mathrm{NFW}(r) r^2 dr = 2m_{\mathrm{BH}}.
\end{equation}
We follow the standard assumption of $r_\mathrm{sp} = 0.2 r_\mathrm{h}$ to determine the spike radius~\cite{merritt2003single,fields2014galactic,eda2015gravitational}. 
The dark matter spike density $\rho_\mathrm{sp}(r)$ follows a power law with spike index $\gamma_{\mathrm{sp}}$ and is given by
\begin{equation}\label{eq:spike}
    \rho_{\mathrm{sp}}(r) = \rho_\mathrm{NFW}(r_{\mathrm{sp}}) \left(\frac{r}{r_{\mathrm{sp}}}\right)^{-\gamma_{\mathrm{sp}}}.
\end{equation}
The spike index $\gamma_{\mathrm{sp}}$ is related to the initial power-law index $\gamma$ of the dark matter density profile of the host halo by~\cite{1995ApJ...440..554Q,gondolo1999dark}
\begin{equation}\label{eq:spike_index}
    \gamma_{\mathrm{sp}} = \dfrac{9-2 \gamma}{4-\gamma}.
\end{equation}
For the NFW profile with $\gamma = 1$ the spike index reduces to $\gamma_{\mathrm{sp}} = 7/3$. The spike density $\rho_\mathrm{sp}(r)$ diverges at small radii, however, the self-annihilation of dark matter particles results in a weak cusp for the dark matter density at the saturation radius $r_{\mathrm{sat}}$ which is given by
\begin{equation}\label{eq:r_sat}
    \rho_{\mathrm{sp}} (r_{\mathrm{sat}}) = \dfrac{m_\chi}{\langle \sigma v \rangle \cdot (t_0 - t_{\mathrm{f}})}
\end{equation}
with the dark matter mass $m_\chi$, the dark matter annihilation cross section times the relative velocity $\langle \sigma v \rangle$, the age of the universe $t_0$ and the formation time of the black hole $t_{\mathrm{f}}$~\cite{gondolo1999dark}.
Therefore, the saturation radius $r_{\mathrm{sat}}$ depends on the assumed dark matter particle. 
For $r \leq r_\mathrm{sat}$ and \textit{s}-wave annihilation, the dark matter distribution follows~\cite{vasiliev2007dark,shapiro2016weak}
\begin{equation}\label{eq:cusp}
    \rho_\mathrm{wcusp}(r) = \rho_\textrm{sp}(r_\mathrm{sat}) \cdot \left( \dfrac{r}{r_\mathrm{sat}} \right)^{-0.5}.
\end{equation}
Finally, we define the cutoff radius $r_\mathrm{cut}$ as
\begin{equation}\label{eq:r_cut}
    r_{\mathrm{cut}} = \mathrm{max}[2r_{\mathrm{schw}}, r_{\mathrm{sat}}].
\end{equation}
For dark matter masses in the GeV to TeV scale and typical dark matter cross sections of $\langle \sigma v \rangle \sim \SI{e-26}{cm^3 s^{-1}}$, the saturation radius $r_{\mathrm{sat}}$ is typically in the order of $\SI{e-3}{\pc}$, so $r_\mathrm{cut}=r_\mathrm{sat}$. \\
A typical dark matter density profile around an IMBH is shown in Figure~\ref{fig:spike_profile}. 
In this particular example, we assume a black hole mass of $\SI{1.6e5}{\Msun}$, the NFW profile as dark matter halo profile with $\rho_0 = \SI{1.9}{\GeV \per \cubic \cm}$ and $r_\mathrm{s} = \SI{2.3}{\kpc}$, a spike radius of $r_\mathrm{sp} = \SI{4.2}{\pc}$ and cutoff radius of $r_\mathrm{cut} = \SI{2.3e-3}{\pc}$. These values correspond to the parameters from a specific and typical IMBH in our analysis, as we show later in Section~\ref{sec:analysis}. \\

\subsection{Gamma-ray flux from dark matter annihilation}
The dark matter density profile around IMBHs summarised above is crucial for predicting the expected gamma-ray flux from dark matter annihilation. As presented by Bertone, Zentner and Silk (2005)~\cite{bertone2005new}, the gamma-ray flux $\Phi$ from dark matter annihilation can be expressed as
\begin{align}\label{eq:flux_general}
    \Phi(E,D) & = \left. \frac{1}{2} \frac{\langle \sigma v \rangle}{m_\chi^2} \frac{1}{D^2} \frac{{\rm d}N}{{\rm d}E} \int_{2r_\mathrm{schw}}^{r_{\rm sp}} \rho^2(r) r^2 dr \right. \\
    & \approx \frac{{\rm d}N}{{\rm d}E} \frac{\langle \sigma v \rangle}{m_\chi^2 D^2} \rho (r_\mathrm{sp})^2 r_\mathrm{sp}^3 \frac{2\gamma_\mathrm{sp} - 1}{8\gamma_\mathrm{sp} - 12} \left(  \frac{r_\mathrm{cut}}{r_\mathrm{sp}} \right)^{3 -2 \gamma_\mathrm{sp}} 
\end{align}
with the dark matter annihilation spectrum $\mathrm{d}N/\mathrm{d}E$ and the distance $D$ from the observer to the IMBH. We assumed $r_\mathrm{cut} \gg r_\mathrm{schw}$ and $r_\mathrm{sp} \gg r_\mathrm{cut}$ to simplify the equation. Unlike the work by Bertone, Zentner and Silk (2005)~\cite{bertone2005new}, we do not neglect the integral for $r<r_\mathrm{cut}$ and take the weak cusp into account.
Therefore, the gamma-ray flux from dark matter annihilation can be calculated for a specific dark matter particle, the distance to the IMBH and its corresponding dark matter spike parameters. 
Assuming $\gamma_{\mathrm{sp}} = 7/3$, we note here that the flux is effectively proportional to $\langle \sigma v \rangle^{2/7}m_\chi^{-9/7}$ since the cutoff radius $r_\mathrm{cut}$ itself is proportional to the dark matter cross section and mass via $r_\mathrm{cut} \propto \langle \sigma v \rangle^{3/7}m_\chi^{-9/7}$ (see Equations~\ref{eq:spike},~\ref{eq:r_sat}, and~\ref{eq:r_cut}).

\section{Dataset}\label{sec:dataset}
\subsection{EAGLE simulations}\label{sec:eagle}
The \textit{Evolution and Assembly of GaLaxies and their Environments} (EAGLE) project~\cite{schaye2015eagle, crain2015eagle} is a cosmological simulation following the evolution of galaxies in a $\Lambda$CDM universe adopting the cosmological parameters advocated by the Planck Collaboration~\cite{ade2014planck}. The simulations are performed using a modified version of the N-Body Tree-PM Smoothed Particle Hydrodynamics (SPH) (GADGET-3) code~\cite{springel2005cosmological}. Gravitationally bound structures are identified using the \textit{Subfind} algorithm~\cite{springel2001populating, dolag2009substructures}. The EAGLE simulations are calibrated to reproduce the stellar mass function, galaxy sizes, and the galaxy mass-black hole mass relation at $z \sim 0$ and include a variety of physical processes, such as star formation and feedback, stellar mass loss, black hole accretion and \textit{active galactic nucleus} (AGN) feedback.
The formation scenarios of black holes ending up in the centre of galaxies, i.e. the remnants of Population III stars, the collapse of cold gas in early-forming and massive halos, or the runaway collisions of stars and stellar mass black holes~\cite{kocsis2014menus}, cannot be resolved by the EAGLE simulations.
Therefore, seed black holes of mass $\SI{e5}{\Msun} /h$ are placed into the centre of halos greater than $\SI{e10}{\Msun} /h$ if they do not already contain a black hole, following the approach in Springel et al. (2005)~\cite{springel2005modelling}. The black holes can grow by accreting gas from their surroundings and by merging with other black holes. Their accretion rate is calculated using the Bondi-Hoyle-Lyttleton accretion rate~\cite{rosas2015impact} and modified as described by Schaye et al. (2015)~\cite{schaye2015eagle}. 
At each time step of the simulation, the black holes are manually repositioned by moving them to the location of the neighbouring particle with the lowest gravitational potential, which has a relative velocity less than $\SI{25}{\percent}$ of the sound velocity and has a distance to the black hole smaller than three gravitational softening lengths.
These conditions ensure that black holes in gas-poor halos do not jump to nearby satellites. \\
The EAGLE simulation used in this work is the reference dataset \code{Ref-L0100N1504} performed in a periodic box with a comoving side length of $L=\SI{100}{cMpc}$ (comoving Mpc), total number of particles of $N=2 \times 1504^3$, initial baryonic particle mass of $m_{\mathrm{g}} = 1.81 \times \SI{e6}{\Msun}$ and total dark matter particle mass of $m_{\chi} = 9.7 \times \SI{e6}{\Msun}$. The comoving Plummer-equivalent gravitational softening length is $\SI{2.66}{ckpc}$ and the maximum physical softening length is $\SI{0.7}{ckpc}$. The database is split into the EAGLE \textit{galaxy database} containing information about halos, galaxies and their merger trees, and the EAGLE \textit{particle data} containing information about each individual gas, dark matter, star and black hole particle within the simulation~\footnote{The data are publicly available at \url{http://icc.dur.ac.uk/Eagle/database.php}.}. This work makes use of both the EAGLE galaxy database and the EAGLE particle data. The galaxy database is used to select Milky Way-like galaxies and to determine the formation halo of the black holes, and the particle data is used to extract information about the black holes themselves, such as their mass and their coordinates. \\

\subsection{Milky Way-like galaxy selection}
The gamma-ray flux from dark matter self-annihilation is antiproportional to the squared distance from the observer to the IMBH, as shown in Equation~\ref{eq:flux_general}. Therefore, we are particularly interested in IMBHs within our own Milky Way, which thus leads to higher gamma-ray fluxes. In the following, we describe how we select galaxies with properties similar to the Milky Way within the EAGLE simulations. \\
We derive our selection criteria of Milky Way-like galaxies using the previous works of Callingham et al. (2019)~\cite{callingham2019mass}, Ortega-Martinez et al. (2022)~\cite{ortega2022milky} and Bignone, Helmi and Tissera (2019)~\cite{bignone2019gaia} as a starting point. Furthermore, based on Wang et al. (2020)~\cite{wang2020mass}, we apply an additional requirement on the halo mass $M_{200}$ defined as the mass enclosed within a sphere of radius $R_{200}$, which is the radius at which the mean density of the halo is 200 times the critical density of the universe. 
Wang et al. (2020) found that the mass $M_{200}$ of the Milky Way is likely to be in the range $0.5-2.0 \times \SI{e12}{\Msun}$ and we therefore select Milky-Way-like galaxies within EAGLE with this particular mass range.
In addition, we require that galaxies have a stellar mass range $M_{*}(r<\SI{30}{kpc})$ of $10^{10.4}$--$10^{11.2}\,\si{\Msun}$ based on Ortega-Martinez et al. (2022)~\cite{ortega2022milky}. The selection criteria regarding the halo mass and stellar mass are close to those used in Sanderson et al. (2020)~\cite{sanderson2020synthetic} based on the FIRE-2 simulation to generate synthetic surveys resembling Gaia DR2 in data structure, magnitude limits, and observational error.
Furthermore, we apply selection cuts on the current star formation rate (SFR) of the galaxy and the stellar disk-to-total mass ratio $f_\mathrm{disk}$. We use the stellar disk-to-total mass ratio $f_\mathrm{disk}$ for massive EAGLE galaxies at redshift $z=0$ from Proctor et al. (2024)~\cite{proctor2024identifying} who applied Gaussian mixture models to the kinematics of stellar particles and identified the disk, bulge, and intra-halo light (IHL) components of EAGLE galaxies. We follow the selection cuts from Bignone, Helmi and Tissera (2019) regarding the SFR of the galaxy and the stellar disk-to-total mass ratio $f_\mathrm{disk}$, i.e. we require the SFR to be in the range $0.1-\SI{3}{\Msun \per \yr}$ and $f_\mathrm{disk} > 0.4$. 
Since the Milky Way has not undergone any major mergers within the past couple of billion years~\cite{ciucua2023chasing}, we also require that the host halos are relaxed systems, i.e. the distance between the centre of mass and the centre of potential of the galaxy is less than $0.07R_{200}$ and that the halo mass in the substructures is less than $\SI{10}{\percent}$ of the halo mass of the galaxy~\cite{callingham2019mass}. Furthermore, we require that the satellite galaxies are located at a distance of $40 \text{ kpc} < r' < 300 \text{ kpc}$, where $r' = r(10^{12} M /M_{200})^{1/3}$ is the distance in units of the virial radius~\cite{callingham2019mass}. This results in satellite distribution similar to the Milky Way. The selection criteria are summarised below. \\ 

\noindent
\textit{Selection criteria for host halos:}
\begin{enumerate}[i)]
    \item Halo mass $M_{200}$ range: $0.5-2.0 \times \SI{e12}{\Msun}$
    \item Stellar mass $M_{*}(r<\SI{30}{kpc})$ range: $10^{10.4}$--$10^{11.2}\,\si{\Msun}$
    \item Star formation rate range: $0.1–\SI{3}{\Msun \per \yr}$
    \item Stellar disk-to-total mass ratio $f_\mathrm{disk}$ larger than 0.4
    \item Distance between the centre of mass and the centre of potential of the galaxy is less than $0.07R_{200}$
    \item Total mass in substructures is less than $\SI{10}{\percent}$ of the total mass of the galaxy
\end{enumerate}

\noindent
\textit{Selection criteria for satellite galaxies:}
\begin{enumerate}[i)]
    \item Distance from halo centre $r$ in the range: $40 \text{ kpc} < r' < 300 \text{ kpc}$ \\
    with $r' = r(10^{12} M_\odot  /M_{200})^{1/3}$
\end{enumerate}

\noindent
We do not consider halos that have been labelled as \textit{spurious} within the EAGLE database since these entities should not be considered as genuine galaxies~\cite{mcalpine2016eagle}.
The resulting dataset consists of about 150 Milky Way-like galaxies and about 6300 associated satellites. In the following, we often refer to the central galaxy in these systems as \textit{main galaxy} and their corresponding satellites as \textit{satellite galaxies}. In the EAGLE simulations, the main galaxies are classified by $\code{SubGroupNumber}=0$ and the satellite galaxies by $\code{SubGroupNumber}>0$.

\section{Analysis}\label{sec:analysis}
\subsection{Properties of selected EAGLE galaxies}\label{sec:analysis_galaxy_properties}
\begin{figure}[htb]
    \centering
  
    \begin{subfigure}[b]{0.325\textwidth}
        \centering
        \includegraphics[width=\textwidth, trim={0.2cm 0 0.2cm 0}, clip]{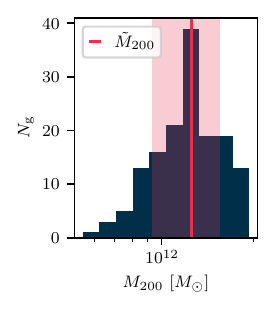}
      \end{subfigure}
    \hfill
    \begin{subfigure}[b]{0.325\textwidth}
        \centering
        \includegraphics[width=\textwidth, trim={0.2cm 0 0.2cm 0}, clip]{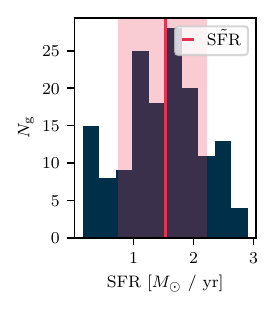}
    \end{subfigure}
    \hfill
    \begin{subfigure}[b]{0.325\textwidth}
        \centering
        \includegraphics[width=\textwidth, trim={0.2cm 0 0.2cm 0}, clip]{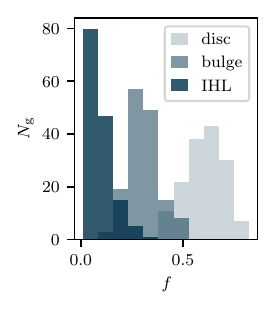}
    \end{subfigure}
  
    \caption{Number of our selection of Milky Way-like galaxies within the EAGLE simulations $N_\mathrm{g}$ versus the galaxy mass $M_{200}$ (left), the star formation rate (SFR) (middle) and the galactic disk, bulge and intra-halo light (IHL) mass fractions $f$ (right). The red lines correspond to the median values and the red shaded regions to the errors on the median, which are calculated by determining the $16^{\mathrm{th}}$ and $84^{\mathrm{th}}$ percentile of the distribution.}
    \label{fig:mw_like}
\end{figure}

We first have a detailed look at our selection of Milky Way-like galaxies within the EAGLE simulations to ensure that the selected main galaxies meet the properties of the Milky Way. Figure~\ref{fig:mw_like} shows the distribution of the main galaxy mass $M_{200}$ (left), the SFR (middle) and the galactic disk, bulge and IHL mass fractions $f$ (right). By construction, the halo mass $M_{200}$ of the our selection of main galaxies ranges between $0.5-2.0 \times \SI{e12}{\Msun}$ with a median halo mass $\tilde{M}_{200}$ of $1.25^{+0.31}_{-0.32} \times \SI{e12}{\Msun}$. Here and in the following sections, the errors on the median are calculated by determining the $16^{\mathrm{th}}$ and $84^{\mathrm{th}}$ percentile of the distribution in order to cover \SI{68}{\percent} of the data around the median. The median halo mass $\tilde{M}_{200}$ agrees very well with the halo mass from a variety of observations using \textit{Gaia DR2} data~\cite{gaia2018vizier,prusti2016gaia} in Wang et al. (2020)~\cite{wang2020mass} which results in $\sim \SI{1.2e12}{\Msun}$. The mass distribution in Figure~\ref{fig:mw_like} (left) shows a distinct peak at $\tilde{M}_{200}$ with significantly less galaxies with masses at the lower and upper end of our mass range. This highlights that the majority of our selected EAGLE galaxies are very close to the current estimate of the halo mass of the Milky Way. The SFR distribution in Figure~\ref{fig:mw_like} (middle) shows a moderate peak around its median star formation rate $\tilde{\mathrm{SFR}}$ of $1.53^{+0.69}_{-0.78}~\SI{}{\Msun \per \yr}$. Chomiuk and Povich (2011)~\cite{chomiuk2011toward} considered different determinations of the SFR in the Milky Way and showed that the SFR converges to $1.9 \pm \SI{0.4}{\Msun \per \yr}$, which lies within the error range of $\tilde{\mathrm{SFR}}$. Lastly, the galactic disk, bulge and IHL mass fraction distributions in Figure~\ref{fig:mw_like} (right) show distinct peaks for each mass fraction. The median values $\tilde{f}_\mathrm{disk}=0.61^{+0.08}_{-0.10}$, $\tilde{f}_\mathrm{bulge}=0.30^{+0.07}_{-0.06}$ and $\tilde{f}_\mathrm{IHL}=0.08^{+0.08}_{-0.03}$ indicate that our selection of main galaxies are composed of a prominent disk component with a smaller bulge and an even smaller IHL component. We note that the stellar disk-to-total mass ratio $f_\mathrm{disk}$ of the Milky Way is measured to be around 0.86~\cite{mcmillan2011mass} and is therefore significantly higher than $f_\mathrm{disk}$ of our selection of EAGLE galaxies. The maximum stellar disk-to-total mass ratio $f_\mathrm{disk}$ of our selection of EAGLE galaxies is about 0.82, meaning that none of the EAGLE galaxies reaches a disk component that is as dominant as the Milky Way disk component. Although this could potentially be a limitation of our analysis, we will show in Section~\ref{sec:correlation_galaxy_prop} that the stellar disk-to-total mass ratio does not seem to have a significant impact on our estimate of the number of IMBHs within our Milky Way. Overall, we conclude that the key properties of our selection of Milky Way-like galaxies within the EAGLE simulations align well with those measured for the Milky Way itself.

\subsection{IMBH number distribution and coordinates}
\begin{figure}[htb]
    \centering
  
    \begin{subfigure}[b]{0.49\textwidth}
        \centering
        \includegraphics[width=\textwidth]{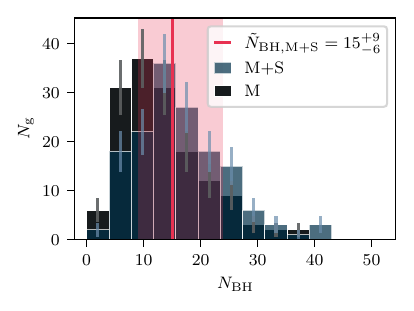}
      \end{subfigure}
    \hfill
    \begin{subfigure}[b]{0.49\textwidth}
        \centering
        \includegraphics[width=\textwidth]{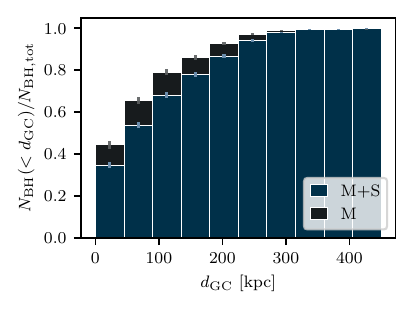}
    \end{subfigure}
  
    \caption{Number distribution (left) and cumulative radial distribution (right) of IMBHs. The distributions for all IMBHs, i.e. IMBHs associated with main or satellite galaxies (M+S), and for the IMBHs associated with the main galaxies only (M) is shown.}
    \label{fig:r_n_dist}
\end{figure}

Given that major mergers of black holes, i.e. mergers of black holes with similar mass, can lead to the disruption of the dark matter spike~\cite{bertone2005new}, we only consider unmerged IMBHs in this analysis. Furthermore, we exclude black holes with a mass $m_\mathrm{BH} > \SI{e6}{\Msun}$ to stay within the mass range of IMBHs. For each Milky-Way like galaxy in the EAGLE simulations, we determine its \code{GroupNumber} and assign IMBHs with the same \code{GroupNumber} to the galaxy. In total, we find about 2500 IMBHs of which about 2000 IMBHs are associated with the main galaxies and the remaining $\sim 500$ IMBHs are associated with satellite galaxies. The number distribution of IMBHs, i.e. the number of IMBHs within a galaxy $N_\mathrm{BH}$ versus the number of galaxies $N_\mathrm{g}$ with $N_\mathrm{BH}$, is shown in Figure~\ref{fig:r_n_dist} (left). We distinguish between the number distribution of all IMBHs, i.e. IMBHs associated with main or satellite galaxies (labelled as 'M+S' in the following), and the number distribution of IMBHs associated with the main galaxies only (labelled as 'M' in the following), thus excluding IMBHs associated with satellite galaxies.
The median number of IMBHs $\tilde{N}_\mathrm{BH, M+S}$ within our selection of galaxies is $15^{+9}_{-6}$, i.e. we expect about 15 IMBHs within a Milky Way-like galaxy and its corresponding satellite galaxies. Considering only IMBHs associated with the main galaxy, we find a median number of IMBHs of $\tilde{N}_\mathrm{BH, M} = 12^{+8}_{-6}$. We further note that $\sim \SI{20}{\percent}$ of satellite galaxies that contain at least one star particle, i.e. excluding dark matter halos, are hosting at least one IMBH. These facts make not only the Milky Way itself an interesting target for IMBHs searches but also underscores the importance of its satellite galaxies. One should notice that the non-observation of IMBHs with masses larger than $\sim \SI{e5}{\Msun}$ in Milky Way satellite galaxies is not in tension with the above-mentioned prediction given the present associated uncertainties.
However, the estimated number of IMBHs within the Milky Way differs significantly from the $101 \pm 22$ IMBHs found by Bertone, Zentner and Silk (2005)~\cite{bertone2005new}. We discuss potential causes for these differences in Section~\ref{sec:discussion}.

\begin{figure}[htbp]
    \centering
    \includegraphics[width=0.99\textwidth, trim=0cm 2cm 0cm 2cm, clip=true]{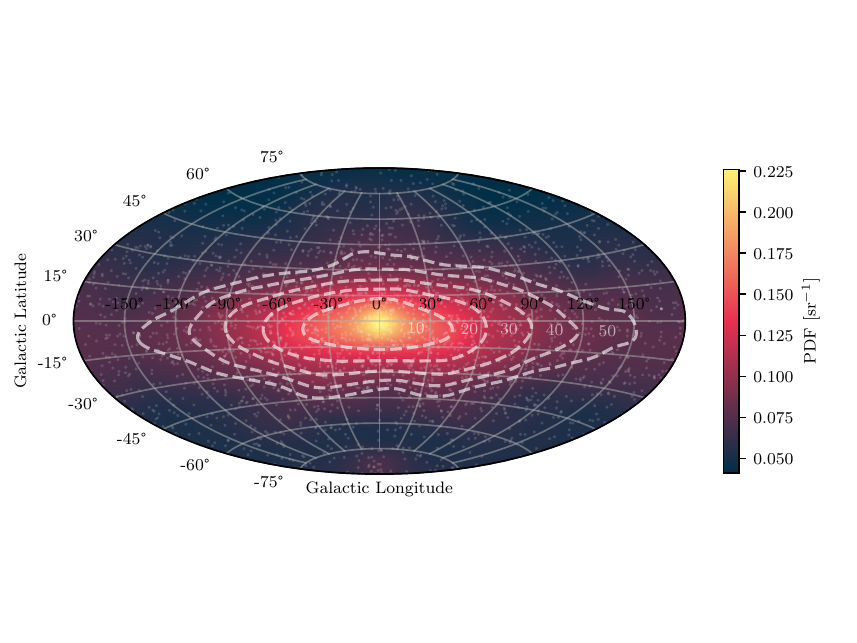}
    \caption{2D map of IMBHs associated with main or satellite galaxies (M+S) in galactic coordinates. The PDF is calculated using a gaussian kernel density estimation and the contours correspond to the integral of the PDF for a given sky area such that they contain $10 \%$, $20 \%$, $30 \%$, $40 \%$ and $50 \%$ of the total number of IMBHs. Only $\SI{1}{\percent}$ of the upsampled IMBH coordinates are depicted here (see text for details).}
    \label{fig:map}
\end{figure}

\noindent
Next, we identify the spatial distribution of IMBHs within their main galaxies, focusing on their radial distribution and their density distribution in galactic coordinates.
For each main galaxy, we determine the coordinates of the IMBHs for a coordinate system with its origin at the galactic centre. We define the galactic centre of each galaxy as the centre of potential, which is given in the EAGLE galaxy database by the \code{CentreOfPotential\_x}, \code{CentreOfPotential\_y} and \code{CentreOfPotential\_z} in the \code{SubHalo} table. The coordinates of the IMBHs are given in the EAGLE particle database by the \code{Coordinates} parameter. As we are interested in the IMBH coordinates in the galactic frame, we rotate the coordinate systems so that the galaxy angular momentum vector (or spin vector), given by \code{Spin\_x}, \code{Spin\_y} and \code{Spin\_z}, is aligned with the $z$-axis of our coordinate system. This step ensures that the disk of the galaxy is located at a galactic latitude of \SI{0}{\degree}. Afterwards, we calculate the distance of the IMBH to the galactic centre $d_{\mathrm{GC}}$ and to the Sun $d_{\mathrm{Sun}}$, and the corresponding galactic coordinates, i.e. galactic latitude $b$ and longitude $l$. We rescale the distance between the Galactic Centre and the Sun $d_\mathrm{GS}'$ based on the halo mass $M_{200}$ of the main galaxy using $d_\mathrm{GS}' = d_\mathrm{GS}(M_{200} / 10^{12} M_\odot)^{1/3}$ with $d_\mathrm{GS} = \SI{8.33}{kpc}$~\cite{cirelli2011pppc}. 
Figure~\ref{fig:r_n_dist} shows the cumulative radial distribution for all IMBHs (M+S) and for IMBHs associated with the main galaxies only (M). In both distributions, the IMBHs are concentrated towards the centre of their main galaxy. As expected, comparing the two cumulative radial distributions indicates that the IMBHs in the satellite galaxies are located at larger distances to the galactic centre. The median distance of all IMBHs to the galactic centre is $\tilde{d}_\mathrm{GC,M+S} = 94^{+124}_{-71} \text{ kpc}$ and the median distance of the IMBHs associated with the main galaxy is $\tilde{d}_\mathrm{GC,M} = 69^{+122}_{-50} \text{ kpc}$. About $80 \%$ of all IMBHs and about $86 \%$ of the IMBHs associated with the main galaxies are within a distance of $\sim \SI{200}{\kpc}$ to the galactic centre.
A 2D map of the positions of IMBHs (M+S) in galactic coordinates is shown in Figure~\ref{fig:map}. For the 2D map, the catalogue was upsampled by randomly generating an angle $\phi_\mathrm{r}$ between 0 and $2\pi$, and adding $\phi_\mathrm{r}$ to the azimuthal angle of the IMBHs in the reference frame with its origin at the galactic centre. The resulting coordinates are added to the IMBH catalogue and the process is repeated 100 times. This way, we achieve an upsampling of the IMBH coordinates by a factor 100 under the assumption that the distribution of IMBHs is independent of the azimuthal angle. The azumithal angle distribution of the original IMBH catalogue was found to be uniformly distributed, which justifies our upsampling method. The \textit{probability density function} (PDF) is calculated using a Gaussian \textit{kernel density estimation} (KDE) with Scott's rule as bandwidth selection method~\cite{scott2015multivariate} and the Haversine metric for distance calculation~\cite{van2012heavenly}. The larger the PDF value in a given region, the higher the IMBH number density in that region. The contours correspond to the integral of the PDF for a given sky area such that they contain $10 \%$, $20 \%$, $30 \%$, $40 \%$ and $50 \%$ of the total number of IMBHs. The PDF shows that the IMBHs are not uniformly distributed in the galaxy. Instead, they are concentrated towards the centre of the galaxy and along the galactic plane. 
We further discuss the consequences of this distribution for the detectability of a dark matter annihilation signal in Section~\ref{sec:discussion}. \\
If not explicitly stated otherwise, we make use of all IMBHs (M+S) for our calculations in the following Sections.

\subsection{IMBH mass and formation redshift}
\begin{figure}[htb]
    \centering
    \begin{subfigure}[b]{0.49\textwidth}
      \centering
      \includegraphics[width=\textwidth]{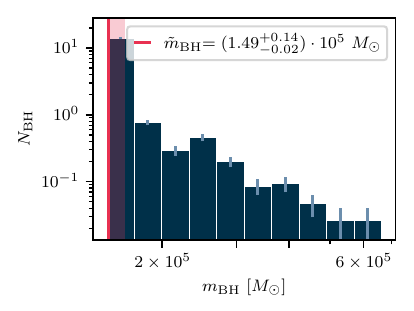}
    \end{subfigure}
    \hfill
    \begin{subfigure}[b]{0.49\textwidth}
      \centering
      \includegraphics[width=\textwidth]{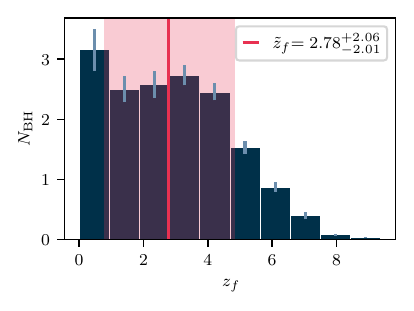}
    \end{subfigure}
    \caption{Mass distribution (left) and formation redshift distribution (right) of IMBHs.}
    \label{fig:m_z_dist}
\end{figure}

In order to calculate the dark matter spike parameters for each individual IMBH, we need to know the mass $m_\mathrm{BH}$ and the formation redshift $z_\mathrm{f}$ of the IMBH. We extract the mass and formation time of the IMBHs at redshift $z=0$ in the EAGLE particle database from the \code{BH\_Mass} and \code{BH\_FormationTime} parameters.
Figure~\ref{fig:m_z_dist} shows the mass and formation redshift distribution of IMBHs within our selection of Milky Way-like galaxies. They are determined by extracting the distribution for each individual galaxy and then calculating the mean of the distributions per bin. Therefore, these distributions represent the average distributions one would expect for a Milky Way-like galaxy. The large majority of IMBHs have a mass close to the initial seeding mass of $\SI{e5}{\Msun} /h = \SI{1.48e5}{\Msun}$. The median mass of IMBHs is $1.49^{+0.14}_{-0.02} \cdot \SI{e5}{\Msun}$ and the distribution rapidly decreases for increasing black hole mass. Since we excluded black holes that encountered any merger during their lifetime, the mass accretion of IMBHs is purely caused by the accretion of gas from their surroundings. \\
The formation redshift distribution of IMBHs is shown in Figure~\ref{fig:m_z_dist} (right). The median formation redshift of IMBHs is $2.78^{+2.06}_{-2.01}$ with the largest number of IMBHs having formed at a low redshift with $z \lesssim 1$. This meets our expectation since the seed black holes are placed into the centre of halos with a total mass larger than $\SI{e10}{\Msun} /h$. The number of halos with a mass larger than $\SI{e10}{\Msun} /h$ increases with increasing evolution time of the universe, resulting in a higher number of IMBHs forming at a lower redshift. \\
Note that the low formation redshift and the small growth of the black holes is a consequence of the limited resolution of the EAGLE simulation. As briefly discussed in Section~\ref{sec:introduction}, IMBHs are expected to form via processes that take place at much higher redshifts, i.e $z \gtrsim 10$~\cite{rosswog2009tidal,loeb1994collapse,karlsson2013pregalactic,kawaguchi2008formation}. However, due to the black hole seeding mechanism applied in EAGLE, most of the black holes in the simulation appear at $z \lesssim 5$. We therefore assume here that the black holes actually formed at higher redshifts and then grew until the seeding mass of $\SI{e5}{\Msun} /h$ was reached. Thus, the majority of the growth of the black holes and consequently the formation of the dark matter spikes takes place before the black holes have reached the EAGLE seeding mass.

\subsection{Correlation between $N_\mathrm{BH}$ and galaxy properties} \label{sec:correlation_galaxy_prop}
\begin{figure}[htb]
    \centering
  
    \begin{subfigure}[b]{0.325\textwidth}
        \centering
        \includegraphics[width=\textwidth, trim={0.2cm 0 0.2cm 0}, clip]{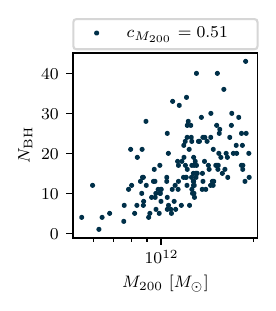}
      \end{subfigure}
    \hfill
    \begin{subfigure}[b]{0.325\textwidth}
        \centering
        \includegraphics[width=\textwidth, trim={0.2cm 0 0.2cm 0}, clip]{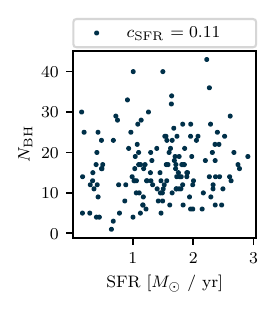}
    \end{subfigure}
    \hfill
    \begin{subfigure}[b]{0.325\textwidth}
        \centering
        \includegraphics[width=\textwidth, trim={0.2cm 0 0.2cm 0}, clip]{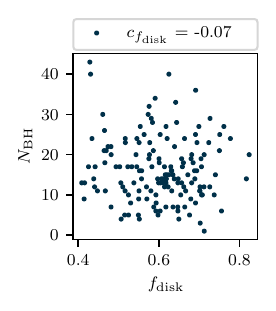}
    \end{subfigure}
  
    \caption{Number of IMBHs $N_\mathrm{BH}$ in each galaxy versus the mass of the galaxy $M_{200}$ (left), the SFR (middle) and the stellar disk-to-total mass ratio $f_\mathrm{disk}$ (right). }
    \label{fig:n_mw_correlation}
\end{figure}

We calculate the correlation between the properties of the main galaxies and the number of IMBHs $N_\mathrm{BH}$ in each galaxy in order to investigate potential indicators for the presence of IMBHs. We calculate the correlation coefficient $c_i$ between the number of IMBHs $N_\mathrm{BH}$ and the galaxy properties as follows:
\begin{equation}
    c_i = \dfrac{\mathrm{cov}(N_\mathrm{BH},i)}{\sigma_{N_\mathrm{BH}} \sigma_i}
\end{equation}
whereas $\mathrm{cov}(X,Y)$ is the covariance of $X$ and $Y$, $\sigma_X$ is the standard deviation of $X$ and $i = \{M_{200}, \mathrm{SFR}, f_\mathrm{disk} \}$. Figure~\ref{fig:n_mw_correlation} shows the number of all IMBH $N_\mathrm{BH}$ in each galaxy versus the mass of the galaxy $M_{200}$ (left), the SFR (middle) and the stellar disk-to-total mass ratio $f_\mathrm{disk}$ (right). The strongest correlation is observed between the number of IMBHs $N_\mathrm{BH}$ and the galaxy mass $M_{200}$ with a correlation coefficient of $c_{M_{200}} = 0.51$. Whereas galaxies with a mass $M_{200} \lesssim \SI{8e11}{\Msun}$ contain no more than $\sim 20$ IMBHs, galaxies with $M_{200} \gtrsim \SI{1.2e12}{\Msun}$ can contain up to $\sim 40$ IMBHs. Both the star formation rate and the stellar disk-to-total mass ratio $f_\mathrm{disk}$ show only a very mild correlation with the number of IMBHs $N_\mathrm{BH}$ with $c_\mathrm{SFR} = 0.11$ and $c_{f_\mathrm{disk}} = -0.07$, respectively. We discuss the consequences of these results in more detail in Section~\ref{sec:discussion}.

\subsection{Dark matter spike parameters}\label{sec:analysis_spike}
\begin{figure}[htb]
    \centering
  
    \begin{subfigure}[b]{0.49\textwidth}
      \centering
      \includegraphics[width=\textwidth]{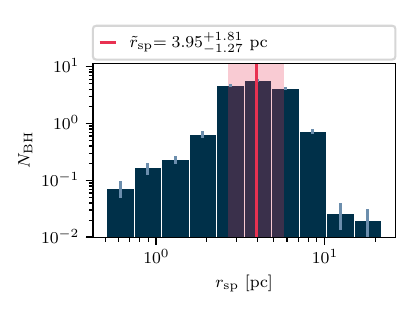}
    \end{subfigure}
    \hfill
    \begin{subfigure}[b]{0.49\textwidth}
      \centering
      \includegraphics[width=\textwidth]{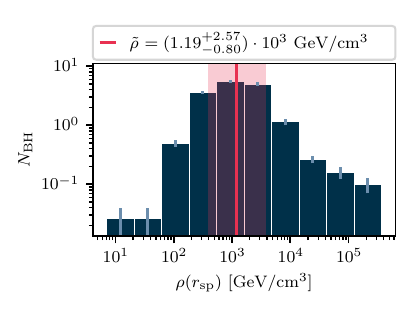}
    \end{subfigure}
  
    \caption{Spike radius distribution (left) and spike density distribution (right)}
    \label{fig:spike_dist}
\end{figure}

For each IMBH, we calculate the dark matter spike parameters $r_{\mathrm{sp}}$, $\rho(r_{\mathrm{sp}})$ and $r_{\mathrm{cut}}$. Therefore, it is necessary to determine the host halo in which the IMBH formed.
We extract the formation redshift $z_\mathrm{f}$ of each IMBH at $z=0$ as described in the previous section and determine the closest redshift $z_\mathrm{c}$ with $z_\mathrm{c} \leq z_\mathrm{f}$ for which a snapshot in the EAGLE simulations is available. In this snapshot, we identify the IMBH based on its \code{ParticleIDs} and identify its host galaxy based on its \code{GroupNumber} and \code{SubGroupNumber}.
We extract the dark matter density profile $\rho_\mathrm{host}(r)$ of the host galaxy using the mass profile $M(<r)$ information, given by the \code{ApertureSize} and \code{Mass\_DM} parameters from the \code{Aperture} table of the EAGLE galaxy database. 
We fit the dark matter density profile to the NFW profile as described in Equation~\ref{eq:nfw} and obtain the normalisation constant $\rho_0$ and the scale radius $r_\mathrm{s}$ using the least squares method~\cite{miller2006method}. We use $\rho_0$ and $r_\mathrm{s}$ to calculate the radius of gravitational influence $r_\mathrm{h}$ using Equation~\ref{eq:r_h}. We apply the IMBH mass at $z=0$ for Equation~\ref{eq:r_h}, assuming an adiabatic growth of the IMBH since its formation time. Finally, the spike radius $r_\mathrm{sp}$ is calculated by $r_\mathrm{sp} = 0.2 r_h$~\cite{merritt2003single} and the spike density $\rho(r_{\mathrm{sp}})$ by evaluating the NFW profile at $r_{\mathrm{sp}}$. \\
In some cases, the IMBH is not assigned to any halo at $z_\mathrm{c}$. In this case, we use the dark matter density profile of the closest halo at $z_\mathrm{c}$ with a mass larger than the required mass to form a IMBH, i.e. $M > 10^{10} M_{\odot}/ h $, to calculate the spike parameters. Therefore, we assume that the IMBH has formed in the next closest halo with sufficient mass. Furthermore, IMBHs assigned to a \textit{spurious} halo at $z_\mathrm{c}$ or containing zero \code{Mass\_DM} values from the \code{Aperture} table are not considered. 
The spike radius and spike density distribution are shown in Figure~\ref{fig:spike_dist}. The median spike radius is $3.95^{+1.81}_{-1.27} \text{ pc}$ and the median spike density is $1.19^{+2.57}_{-0.80} \cdot \SI{e3}{\GeV \per \cubic \cm}$. 

\section{Results}\label{sec:results}
\begin{figure}[ht]
    \centering
    \begin{subfigure}[b]{0.49\textwidth}
      \centering
      \includegraphics[width=\textwidth]{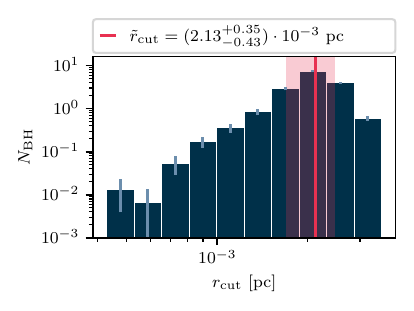}
    \end{subfigure}
    \caption{Cutoff radius distribution for $m_{\chi}=\SI{500}{\GeV}$, $\langle \sigma v \rangle = \SI{3e-26}{\cubic \cm \per \s}$ and the $b \overline{b}-$annihilation channel.}
    \label{fig:r_cut}
\end{figure}

\begin{figure}[htb]
    \centering
  
    \begin{subfigure}[b]{0.49\textwidth}
      \centering
      \includegraphics[width=\textwidth]{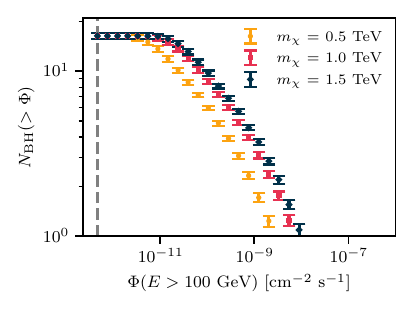}
    \end{subfigure}
    \hfill
    \begin{subfigure}[b]{0.49\textwidth}
      \centering
      \includegraphics[width=\textwidth]{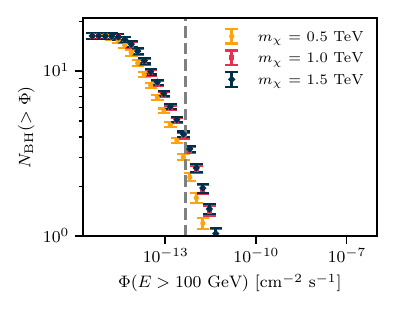}
    \end{subfigure}
  
    \caption{Integrated luminosity of IMBHs for $\langle \sigma v \rangle = \SI{3e-26}{\cubic \cm \per \s}$ (left) and $\langle \sigma v \rangle = \SI{7e-37}{\cubic \cm \per \s}$ (right) for dark matter masses between $\SI{0.5}{\TeV}$ and $\SI{1.5}{\TeV}$, the $b \overline{b}$-channel and $E_\mathrm{th} = \SI{100}{\GeV}$. The grey dashed line depicts the average H.E.S.S. flux sensitivity for the H.E.S.S. Galactic Plane survey~\cite{abdalla2018hess}.}
    \label{fig:int_lum}
\end{figure}

\begin{figure}[htb]
    \centering
  
    \begin{subfigure}[b]{0.49\textwidth}
      \centering
      \includegraphics[width=\textwidth]{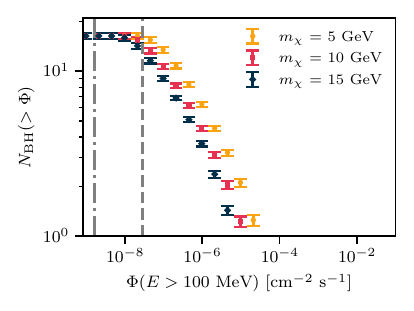}
    \end{subfigure}
    \hfill
    \begin{subfigure}[b]{0.49\textwidth}
      \centering
      \includegraphics[width=\textwidth]{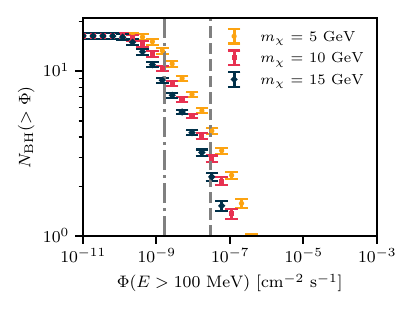}
    \end{subfigure}
  
    \caption{Integrated luminosity of IMBHs for $\langle \sigma v \rangle = \SI{3e-26}{\cubic \cm \per \s}$ (left) and $\langle \sigma v \rangle = \SI{e-32}{\cubic \cm \per \s}$ (right) for dark matter masses between $\SI{5}{\GeV}$ and $\SI{15}{\GeV}$, the $\tau^- \tau^+$-channel and $E_\mathrm{th} = \SI{100}{\MeV}$. The grey dashed (dash-dotted) line depicts the Fermi flux sensitivity for galactic $l = \SI{0}{\deg}$ and $b = \SI{0}{\deg}$ ($l = \SI{120}{\deg}$ and $b = \SI{45}{\deg}$)~\cite{fermi_sensitivity}.}
    \label{fig:int_lum_fermi}
\end{figure}

In this section, we present our results for the detectability of a gamma-ray signal from dark matter annihilation around IMBHs in the Milky Way. Assuming a dark matter mass $m_\chi$ and cross section $\langle \sigma v \rangle$, we calculate the cutoff radius $r_{\mathrm{cut}}$ using Equation~\ref{eq:r_cut}. Figure~\ref{fig:r_cut} shows the distribution of the cutoff radius for $m_\chi = \SI{500}{\GeV}$ and $\langle \sigma v \rangle = \SI{3e-26}{\cubic \cm \per \s}$, assuming the $b \overline{b}-$annihilation channel. The median cutoff radius is $2.13^{+0.35}_{-0.43} \cdot \SI{e-3}{\pc}$.
The larger the assumed dark matter cross section $\langle \sigma v \rangle$, the larger the cutoff radius as more self-annihilation events occur and the saturation of the dark matter spike is reached at larger radii. 
\\
The expected gamma-ray flux from dark matter annihilation is calculated by implementing the distance $D$ to the IMBH, the dark matter mass $m_\chi$, cross section $\langle \sigma v \rangle$ and spike parameters $r_{\mathrm{sp}}$, $\rho(r_{\mathrm{sp}})$ and $r_{\mathrm{cut}}$ into Equation~\ref{eq:flux_general}. We compute the integrated luminosity of IMBHs by calculating the number of IMBHs $N_\mathrm{BH}(\Phi)$ that surpass a certain flux threshold $\Phi(E>E_\mathrm{th})$, assuming a typical energy threshold of (a) $E_\mathrm{th} = \SI{100}{\GeV}$ for Imaging Atmospheric Cherenkov Telescopes (IACTs) and (b) $E_\mathrm{th} = \SI{100}{\MeV}$ for space-based gamma-ray observatories. Figure~\ref{fig:int_lum} (left) shows the average integrated luminosity over all Milky Way-like galaxies for dark matter masses between $\SI{0.5}{\TeV}$ and $\SI{1.5}{\TeV}$, fixed annihilation cross section of $\langle \sigma v \rangle = \SI{3e-26}{\cubic \cm \per \s}$, the $b \overline{b}$-channel and $E_\mathrm{th} = \SI{100}{\GeV}$. For the range of dark matter masses and energy threshold chosen here, the integrated luminosity increases for increasing dark matter mass. This is due to the dark matter annihilation spectrum being integrated from $E_\mathrm{th}$ to $m_\chi$, resulting in an increasing number of photons for higher dark matter mass. The average H.E.S.S. flux sensitivity for the H.E.S.S. galactic plane survey is $\sim \SI{5e-13}{\per \square \cm \per \s}$~\cite{abdalla2018hess} and depicted by the grey vertical line in Figure~\ref{fig:int_lum}. 
Independent of the dark matter masses considered here, all IMBH fluxes are expected to surpass the H.E.S.S. sensitivity, making IMBHs promising targets for ground-based gamma-ray observatories. In order to test the limits of our analysis, we lowered the dark matter cross section until only $\sim 2.3$ IMBHs would exceed the H.E.S.S. sensitivity. This number is motivated by the $\SI{90}{\percent}$ confidence level of a non-detection from Poisson statistics. We find that $\sim 2.3$ IMBHs exceed the H.E.S.S. sensitivity for a dark matter mass of $m_\chi = \SI{500}{\GeV}$ at a cross section of $\sim \langle \sigma v \rangle = \SI{7e-37}{\cubic \cm \per \s}$, see Figure~\ref{fig:int_lum} (right). However, this cross section cannot be directly translated to an upper cross section limit that H.E.S.S. would be able to probe because (a) H.E.S.S. does not have a full sky coverage and (b) H.E.S.S. does not reach the flux sensitivity of the galactic plane survey in all of its observations. We discuss these limitations in more detail in the next section. Figure~\ref{fig:int_lum_fermi} (left) shows the average integrated luminosity for dark matter masses between $\SI{5}{\GeV}$ and $\SI{15}{\GeV}$, fixed annihilation cross section of $\langle \sigma v \rangle = \SI{3e-26}{\cubic \cm \per \s}$, the $\tau^- \tau^+$-channel and $E_\mathrm{th} = \SI{100}{\MeV}$. The 10-years Fermi-LAT flux sensitivity for two different sky positions, $l = \SI{0}{\deg}$ \& $b = \SI{0}{\deg}$, and $l = \SI{120}{\deg}$ \& $b = \SI{45}{\deg}$, are depicted by the dashed and dashed-dotted lines, respectively~\cite{fermi_sensitivity}. The Fermi-LAT flux sensitivity at the Galactic Centre is lower than at higher galactic latitudes and longitudes due to the higher gamma-ray background from the diffuse galactic emission.
All IMBH fluxes are expected to surpass the Fermi-LAT sensitivity at large galactic longitudes and latitudes independent of the dark masses chosen here. Additionally, the majority of IMBH fluxes are expected to surpass the Fermi-LAT Galactic Centre sensitivity. The fact that Fermi-LAT's flux sensitivity is high enough for both small and large galactic coordinates to detect the majority of IMBHs indicates that the instrument has the potential in detecting a gamma-ray signal from dark matter annihilation around IMBHs independent of the IMBH sky position. Similarly to Figure~\ref{fig:int_lum} (right), we lower the dark matter cross section until $\sim 2.3$ IMBHs would exceed the Fermi-LAT flux sensitivity at the Galactic Centre. We find this to be the case at $\langle \sigma v \rangle \sim \SI{e-32}{\cubic \cm \per \s}$. The corresponding integrated luminosity is shown in Figure~\ref{fig:int_lum_fermi} (right). Unlike ground-based gamma-ray observatories, Fermi-LAT provides data for the full gamma-ray sky and is therefore not limited to a specific sky region, such as the Galactic Plane. We discuss the consequences of this fact in more detail in the next section. \\
Furthermore, we investigate the impact of different dark matter density profiles on our results. In addition to the NFW profile, we assume that the dark matter density of the IMBH formation halos follows a cored density profile. Since we find that the EAGLE data does not provide a sufficient resolution in order to properly test the cored profile, we refer the interested reader to the results in Appendix~\ref{sec:cored}. A comparison of different dark matter annihilation channels and their impact on the expected gammm-ray flux around IMBHs is discussed in Appendix~\ref{sec:annihilation_channels}.

\section{Discussion}\label{sec:discussion}
\subsection{Number of IMBHs}
Our analysis provides the number and distribution of IMBHs in a Milky Way-like galaxy under the assumption that a $\SI{e10}{\Msun} /h$ halo is populated by one IMBH with $\SI{e5}{\Msun} /h$ that subsequently grows through accretion. This black hole formation scenario in the EAGLE simulations is motivated by two key factors: firstly, the resolution of the EAGLE simulations is insufficient to accurately represent the actual formation processes of black holes. Secondly, the feedback resulting from the growth of these black holes plays a pivotal role in the formation of galaxies, influencing star formation in massive galaxies and altering the gas profiles of their host halos~\cite{schaye2015eagle,crain2015eagle}.
In this minimal scenario for IMBH formation, the average number of IMBHs within a Milky Way-like galaxy and its corresponding satellite galaxies is $15^{+9}_{-6}$, which is significantly lower than the $101 \pm 22$  IMBHs found by Bertone, Zentner and Silk (2005)~\cite{bertone2005new}. This difference is likely caused by the different black hole seeding mechanisms considered in our analysis. The EAGLE simulations seed a black hole in halos with masses greater than $\SI{e10}{\Msun}/h$. On the other hand, the work of Bertone, Zentner and Silk (2005) follows the procedure of Koushiappas, Bullock, and Dekel (2004)~\cite{koushiappas2004massive}, in which the seeding of black holes is based on a critical halo mass as a function of, among others, the redshift and gas temperature. This leads to host halo masses down to $\sim \SI{e7}{\Msun}$, which allows black holes to form in halos up to three orders of magnitudes smaller than in our analysis. The effect is indirectly illustrated in the redshift distribution shown in Figure~\ref{fig:m_z_dist} (right). Whereas the majority of the black holes in our analysis are seeded at low redshifts with $z \lesssim 5$, the black holes in Bertone, Zentner and Silk (2005) are seeded at redshift $z \gtrsim 10$. This is due the fact that the halos acquire more mass over time and reach the required (lower) seeding mass sooner in the work of Bertone, Zentner and Silk (2005). 
Despite these technical arguments, the approach we employ with the EAGLE simulations offers a more timely understanding of the formation and evolution of galaxies and is validated against the latest observations. While the Bertone, Zentner and Silk (2005) includes smaller progenitor halos in their semi-analytical models, these models lacked the dynamic environmental effects and feedback mechanisms that are now known to significantly impact galaxy formation and black hole growth. EAGLE, on the other hand, represent state-of-the-art cosmological simulations that have been calibrated against a range of observables in our Universe. These simulations include, among others, detailed modelling of star formation, stellar evolution, metal enrichment, supernova feedback, and AGN feedback. These processes are crucial as they influence the thermodynamic properties of the interstellar and intergalactic medium, and hence the formation and evolution of galaxies and black holes within these environments.
Finally, while we are confident that our results represent a more robust estimate based on our current understanding of galaxy evolution, we acknowledge that they are not the only possible predictions. The seeding mechanism used in EAGLE is supported by a wide range of observations~\cite{crain2015eagle} but does not exclude other plausible scenarios or models. \\
Furthermore, it is important to note that the number of IMBHs $N_\mathrm{BH}$ within our selection of Milky Way-like galaxies scatters significantly between the individual galaxies, varying from a minimum of 1 IMBH per galaxy up to a maximum of 43 IMBHs per galaxy. This number strongly correlates with the galaxy mass $M_{200}$ as can be seen in Figure~\ref{fig:n_mw_correlation} (left). Other mass-related parameters, such as the total or star mass of the galaxy show a similar correlation with the number of IMBHs. A precise measurement of $M_{200}$ of the Milky Way is rather challenging and different approaches can lead to significantly different results, see Wang et al. (2020)~\cite{wang2020mass} for a detailed comparison. Our choice of the $M_{200}$-range to select Milky Way-like galaxies within EAGLE is well motivated and aligns with the range of the actual $M_{200}$ measurements of the Milky Way but future, more precise constrains of the Milky Way mass will improve our predictions of the number of IMBHs and make them more robust. The actual stellar disk-to-total mass ratio $f_\mathrm{disk}$ of the Milky Way is measured to be around 0.86, which is significantly higher than $f_\mathrm{disk}$ of most of our selected EAGLE galaxies. However, as shown in Figure~\ref{fig:n_mw_correlation} (right), the correlation between $f_\mathrm{disk}$ and the number of IMBHs is very mild with a correlation coefficient of $c_{f_\mathrm{disk}} = -0.07$. The number of IMBHs within the Milky Way predicted by our analysis is therefore expected to be only mildly affected by the difference between $f_\mathrm{disk}$ of our selected EAGLE galaxies and the true $f_\mathrm{disk}$ value of the Milky Way. Lastly, we do not find a strong correlation between $N_\mathrm{BH}$ and the SFR with a correlation coefficient of $c_\mathrm{SFR} = 0.11$ as shown in Figure~\ref{fig:n_mw_correlation} (middle). This indicates that the presence of IMBHs barely promotes the total SFR of our selection of EAGLE galaxies. 

\subsection{Spatial distribution of IMBHs}
We find that the IMBHs are not uniformly distributed in the galaxy, but that they are concentrated towards the centre of the galaxy and along the Galactic Plane. We suppose that the IMBHs do not follow a uniform distribution due to the manual repositioning of black holes applied in the EAGLE simulations. Although the positions and trajectories of massive black holes are affected by dynamical friction in reality, EAGLE lacks the resolution to capture this effect. Instead, the effect of dynamical friction is modelled by manually repositioning black holes to the location of the neighbouring particle with the lowest gravitational potential at each time step of the simulation~\cite{schaye2015eagle}, as already discussed in Section~\ref{sec:eagle}. Without repositioning, black hole growth is negligible and SMBHs do not end up in the centre of massive galaxies~\cite{bahe2022importance} which would be in strong contradiction with observations. The manual repositioning of black holes applied in the EAGLE simulations therefore seems to capture the effect of dynamical friction reasonably well and is considered to be the main explanation for the IMBH distribution along the Galactic Plane. For a detailed study of the effect of black hole repositioning in galaxy formation simulations, we refer the reader to Bahé et al. (2022)~\cite{bahe2022importance}. Additionally, we suppose that the rather late seeding of the black holes at low redshifts is further contributing to the spatial distribution of the black holes along the Galactic Plane. Due to their injection at low redshifts, the black holes are more prone to interaction with the Galactic Plane which therefore makes them more spatially correlated to the baryonic matter in the Galactic Plane. \\
Furthermore, we observe that the majority of the IMBHs are located within the main galaxy, although a considerable number is also found in the associated satellite galaxies. On average $15^{+9}_{-6}$ IMBHs are distributed within the main galaxy and its corresponding satellites, and $12^{+8}_{-6}$ IMBHs are present in the main galaxy only, indicating that $3^{+2}_{-2}$ of the $15^{+9}_{-6}$ IMBHs are distributed within satellite galaxies. It is also interesting to note that, within our selection of Milky Way-like galaxies, about $\SI{20}{\percent}$ of the satellite galaxies with at least one star particle contain at least one IMBH. For the Milky Way, more than 60 satellite galaxies within $\sim \SI{400}{\kpc}$ have been observed so far~\cite{drlica2020milky}. This makes not only the Milky Way itself but also its satellite galaxies an interesting target for IMBH searches. \\
Given the distribution of IMBHs within both the main galaxy and its satellite galaxies, ground-based gamma-ray observatories, which do not have full sky coverage, should therefore focus on observations of the Galactic Centre, the Galactic Plane, and satellite galaxies to maximise the number of IMBHs within their field of view.
Fortunately, all current generation ground-based gamma-ray observatories, i.e. H.E.S.S., MAGIC and VERITAS, have a Galactic Centre survey and observations of many satellite galaxies in their program~\cite{abdalla2018hess,abdalla2022search,albert2006observation,archer2014very,abramowski2014search,archambault2017dark,magic2016limits}. Due to the high gamma-ray fluxes that are expected from the dark matter annihilation around IMBHs, we do not expect the flux sensitivity to be the limiting factor for the detectibility of a potential gamma-ray signal with ground-based observatories. Instead, the low expected number of IMBHs within the Milky Way is likely to limit the detectibility of a potential gamma-ray signal. Integrating the PDF from Figure~\ref{fig:map} over the sky region of the H.E.S.S. galactic plane survey~\cite{abdalla2018hess}, i.e. for $ \SI{-65}{\degree} < l < \SI{110}{\degree}$ and $|b| < \SI{3}{\degree}$, and excluding the IMBHs within satellite galaxies, we expect about $N_{\mathrm{BH, HESS}} = 0.6^{+0.4}_{-0.3}$ IMBHs within the field of view. The upcoming Cherenkov Telescope Array Observatory (CTAO)~\cite{acharya2017science} is planning to observe the galactic plane covering a larger sky area with $|l| < \SI{90}{\degree}$ and $|b| < \SI{6}{\degree}$, resulting in an expected number of IMBHs within the field of view of $N_{\mathrm{BH, CTAO}} = 1.1^{+0.8}_{-0.6}$. The H.E.S.S. extragalactic survey (HEGS)~\cite{hegs2024} covering a set of extragalactic observations will increase the sky area covered by H.E.S.S. and therefore improve the chances of detecting an IMBH signal although the average flux sensitivity is not expected to be as high as for the galactic plane survey. Other gamma-ray observatories, such as HAWC~\cite{abeysekara2013sensitivity}, LHAASO~\cite{di2016lhaaso} and Fermi-LAT~\cite{atwood2009large}, are able to observe (almost) the full sky and are therefore very well suited for the detection of a gamma-ray signal from dark matter annihilation around IMBHs. 

\subsection{Gamma-ray flux detectability of IMBHs}
The integrated luminosity functions of IMBHs shown in Figure~\ref{fig:int_lum} and Figure~\ref{fig:int_lum_fermi} indicate that a potential gamma-ray signal from dark matter annihilation around IMBHs is detectable with current and future gamma-ray observatories. These objects are expected to appear as unidentified, point-like sources with identical energy spectra. Depending on the experiment, such analyses should be able to probe a wide range of dark matter masses, ranging from sub-GeV to tens of TeV, and cross sections down to $\langle \sigma v \rangle \sim \SI{7e-37}{\cubic \cm \per \s}$ in the most optimistic scenario. To our knowledge, these limits would be the most stringent limits on dark matter annihilation cross sections for dark matter masses in the range of $\sim \SI{}{\GeV} - \SI{}{\TeV}$. However, we emphasize that the search for dark matter spikes around IMBHs is affected by significant uncertainties. The predicted number and spatial distribution of IMBHs in the Milky Way depends on the assumed formation scenario and seeding mechanism applied within a simulation~\cite{bertone2009dark}. Furthermore, our limits on the dark matter cross section depend on the expected number of IMBHs within a galaxy, which is strongly correlated to the mass $M_{200}$ of the galaxy. This introduces an additional systematic uncertainty that can only be reduced by determining the mass of the Milky Way more precisely in the future. Moreover, the IMBH formation scenarios have an impact on the dark matter distribution around the IMBHs at $z=0$. The formation of IMBHs at high redshift is still subject of current research and the details remain to be fully understood~\cite{greene2020intermediate}. The dark matter cross section an analysis is able to probe is directly influenced by those uncertainties. Deriving strict limits associated with high confidence levels is therefore challenging and the limits provided in this article should be treated with caution. However, by integrating state-of-the-art cosmological simulations and implementing recent measurements of the Milky Way, we argue that our analysis offers a novel and greatly improved approach on identifying dark matter spikes around IMBHs in comparison to previous studies.
Here, we have covered the case of dark matter annihilation into gamma rays, but the analysis can be easily extended to other indirect detection channels, such as the detection of neutrinos~\cite{bertone2006prospects} with the IceCube~\cite{abbasi2012design} and KM3NeT~\cite{adrian2016letter} experiments, using the IMBH catalogue and the source code provided in this work. Furthermore, IMBHs could potentially emit X-ray and radio emissions from the accretion of matter in the accretion disk. Gaggero et al. (2017)~\cite{gaggero2017searching} and Scarcella et al. (2021)~\cite{scarcella2021multiwavelength} investigated the multi-wavelength detectability of primordial and astrophysical black holes, respectively. They found that black holes concentrated in the denser central regions of the Galaxy are more likely to accrete gas and produce detectable emissions compared to those located in the less dense outer regions. However, the high gas density region along the Galactic Disk spans only a few $\sim \SI{0.1}{\deg}$ in Galactic latitude. Although we find that the IMBHs in our analysis are mainly distributed towards the Galactic Centre and along the Galactic Disk, the concentration of these objects within $|b| \lesssim \SI{0.3}{\deg}$ is very low. It is therefore unlikely to find an IMBH within this region and we do not expect the IMBHs from our analysis to be strong multiwavelength emitters. We postpone a more detailed analysis of the multiwavelength detectability of these objects to future work. \\
Our IMBH catalogue and the catalogue of our selection of Milky Way-like galaxies within EAGLE is publicly available at~\cite{aschersleben_2024_10491705}. The galaxy catalogue contains, among others, the mass parameters, the SFR and the stellar disk-to-total mass ratio of each individual galaxy. The IMBH catalogue contains, among others, the coordinates, mass, formation redshift and spike parameters for each individual IMBH. Each column of the catalogues is described in detail in Table~\ref{tab:galaxy} $\&$~\ref{tab:imbh}. We also provide separate files for which we calculated the gamma-ray fluxes for different dark matter masses $m_{\chi}$ and annihilation cross sections $\langle \sigma v \rangle$. The columns of these files are described in Table~\ref{tab:fluxes}. The source code used to generate the IMBH catalogue and the gamma-ray fluxes is publicly available at~\cite{aschersleben_2024_10566372}. It provides a detailed description of the analysis steps and can be used to generate the IMBH catalogue for different EAGLE simulations.

\section{Conclusions \& Future Work}\label{sec:conclusion}
We presented a mock catalogue of IMBHs and their dark matter spikes in Milky Way-like galaxies derived from the EAGLE simulations. The catalogue contains the coordinates, mass, formation redshift and dark matter spike parameters for each individual IMBH. On average, our selection of Milky Way-like galaxies contains $15^{+9}_{-6}$ IMBHs, primarily distributed towards the Galactic Centre and along the Galactic Plane. We demonstrated that the gamma-ray flux from dark matter annihilation around IMBHs should be detectable by both current and forthcoming gamma-ray observatories, including H.E.S.S, Fermi-LAT and the upcoming Cherenkov Telescope Array Observatory (CTAO). Depending on the experiment, such analyses should be able to probe a wide range of dark matter masses, ranging from sub-GeV to tens of TeV, and cross sections down to $\langle \sigma v \rangle \sim \SI{7e-37}{\cubic \cm \per \s}$ in the most optimistic scenario. To the best of our knowledge, these limits would be the most stringent constraints on dark matter annihilation cross sections for dark matter masses in the range of $\sim \SI{}{\GeV} - \SI{}{\TeV}$. However, it is crucial to consider the limitations of the EAGLE simulations in capturing the complex processes involved in IMBH formation due to their spatial and temporal resolutions, as highlighted by recent studies of IMBH formation~\cite{fujii2024simulations,giersz2015mocca,fragione2022repeated,rizzuto2021intermediate}. These limitations introduce systematic uncertainties to our analysis that may have a substantial impact on our results. The IMBH catalogue and the source code used to generate the catalogue and calculate the gamma-ray fluxes are publicly available at~\cite{aschersleben_2024_10566372} and~\cite{aschersleben_2024_10491705}. The source code provides a detailed description of the analysis steps and can be used to generate the IMBH catalogue for different EAGLE simulations. In future work, we aim to apply to the IMBH mock catalogue on data of the H.E.S.S. Galactic Plane survey, the H.E.S.S. extragalactic survey and on observations of satellite galaxies to search for a potential gamma-ray signal from dark matter annihilation around IMBHs by investigating the unidentified point sources within these observations. If none of the sources matches the expected dark matter annihilation spectrum, we will provide upper limits on the dark matter cross section.

\acknowledgments
We extend our gratitude to Maxime Trebitsch and Tom Callingham for their assistance and insightful discussions regarding the EAGLE simulations and the Milky Way-like selection criteria. Their contributions have greatly enriched our research endeavors. We thank Katy Proctor and her collaborators for sharing with us the stellar mass ratios of the EAGLE galaxies and for their underlying great work. Dieter Horns acknowledges support by the Deutsche Forschungsgemeinschaft (DFG, German Research Foundation) under Germany's Excellence Strategy - EXC 2121 'Quantum Universe' - 390833306.


\bibliographystyle{JHEP}
\bibliography{biblio.bib}
\clearpage

\appendix
\section{Cored dark matter density profile}\label{sec:cored}
\begin{figure}[htb]
    \centering
  
    \begin{subfigure}[b]{0.49\textwidth}
      \centering
      \includegraphics[width=\textwidth]{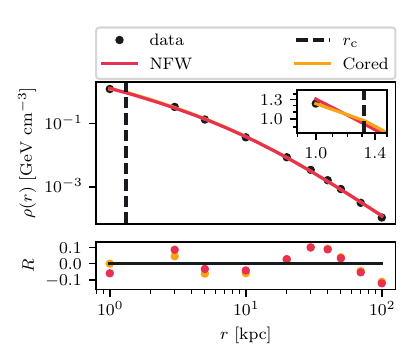}
    \end{subfigure}
    \hfill
    \begin{subfigure}[b]{0.49\textwidth}
      \centering
      \includegraphics[width=\textwidth]{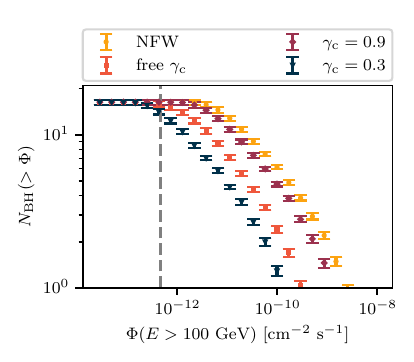}
    \end{subfigure}
  
    \caption{Left: Example of a dark matter halo profile obtained from the EAGLE simulations. The red line represents the best fit model for the NFW profile and the yellow line the best fit model for the cored profile in which the core index $\gamma_\mathrm{c}$ is a free parameter. The inset plot shows a zoom of the $\sim \SI{1}{\kpc}$ region. The bottom panel shows the residuals $R$ for both models. Right: Integrated luminosity of IMBHs obtained under the assumption of a NFW profile, a cored profile with $\gamma_\mathrm{c}$ as free parameter, a cored profile with fixed $\gamma_\mathrm{c}=0.9$ and a cored profile with fixed $\gamma_\mathrm{c}=0.3$. A dark matter particle with $\langle \sigma v \rangle = \SI{3e-26}{\cubic \cm \per \s}$ and $m_\chi = \SI{500}{\GeV}$, and the $b\overline{b}$-channel was assumed. The grey dashed line depictes the average H.E.S.S. flux sensitivity for the H.E.S.S. Galactic Plane survey~\cite{abdalla2018hess}.}
    \label{fig:cored}
\end{figure}

In this section, we examine the impact of a cored dark matter density profile on the gamma-ray flux resulting from dark matter annihilation around IMBHs. The precise shape of the dark matter density profile of galaxies remains a subject of debate. While some studies favour a cusp dark matter density profile, such as the NFW profile~\cite{navarro1997universal}, others advocate a cored dark matter density profile~\cite{sameie2021central}. Therefore, we investigate the impact of different dark matter profile models on our results. In addition to the NFW profile defined in Equation~\ref{eq:nfw}, we introduce a cored dark matter profile as
\begin{align}
    \label{eq:cored_density}
        \rho_\mathrm{cored}(r)= \left\{
            \begin{array}{ll}
                \displaystyle\rho_\textrm{NFW}(r_\mathrm{c}) \left( \dfrac{r}{r_\mathrm{c}}\right)^{-\gamma_\mathrm{c}} & \quad r < r_\mathrm{c} \\
                \rho_\textrm{NFW}(r) & \quad r \geq r_\textrm{c}
            \end{array}
        \right.
\end{align}
characterized by the core index $\gamma_\mathrm{c}$ (where $0 \leq \gamma_\mathrm{c} < 1$) and the core radius $r_\mathrm{c}$, which is expected to be in the order of $\sim \SI{1}{\kpc}$~\cite{sameie2021central}. Following the same approach as presented in Section~\ref{sec:analysis_spike}, we fit the dark matter density profile of the IMBH formation galaxy to the cored dark matter density profile and obtain the best fit values for $\rho_0$, $r_\mathrm{s}$, $\gamma_\mathrm{c}$ and $r_\mathrm{c}$. Figure~\ref{fig:cored} (left) shows an example of a dark matter density profile of a IMBH formation galaxy obtained from the EAGLE simulations and the best fit models for the NFW profile and the cored profile. The bottom panel shows the residuals $R$, i.e. $(data-model)/model$, of both models and the inset plot provides a zoom in of the $\sim \SI{1}{\kpc}$ region. In this particular example, the best fit parameters are $\gamma_\mathrm{c} = 0.82$ and $r_\mathrm{c} = 1.32 \pm \SI{0.32}{\kpc}$. The error on the core index $\gamma_\mathrm{c}$ is negligible. Both models describe the dark matter density profile very well with only minor differences between the two models at small radii. However, given that the EAGLE data provides merely one data point for $r \lesssim \SI{1}{\kpc}$, an accurate fit for the cored profile in this region is not expected. Therefore, the best fit parameters $\gamma_\mathrm{c}$ and $r_\mathrm{c}$ should be treated with great care. In order to accurately determine a cored dark matter profile for the host galaxies, more data for $\sim \SI{1}{\kpc}$ would be needed. Nevertheless, we proceed with our approach to further explore the effects of varying core indices on the resulting gamma-ray fluxes. Therefore, we calculate the dark matter spike parameters, i.e. $r_\mathrm{sp}$, $\rho(r_\mathrm{sp})$, $r_\mathrm{cut}$ and $\gamma_\mathrm{sp}$, using the same approach as described in Section~\ref{sec:analysis_spike}. We investigate three different scenarios for the core index $\gamma_\mathrm{c}$: (i) we vary the core index $\gamma_\mathrm{c}$ during the fit, (ii) we fix the core index to $\gamma_\mathrm{c}=0.9$ and (iii) we fix the core index to $\gamma_\mathrm{c}=0.3$. Our choice of $\gamma_\mathrm{c}=0.9$ and $\gamma_\mathrm{c}=0.3$ is motivated by findings with the FIRE simulation of baryons and cold dark matter~\cite{sameie2021central}. Figure~\ref{fig:cored} (right) shows the integrated luminosity function of IMBHs assuming the NFW profile and the cored profile for $\gamma_\mathrm{c}$ as a free fitting parameter, $\gamma_\mathrm{c}=0.9$ and $\gamma_\mathrm{c}=0.3$. The cored profile consistently yields lower gamma-ray fluxes compared to the NFW profile, regardless of our varied approaches for $\gamma_\mathrm{c}$. The core index $\gamma_\mathrm{c}$ as free parameter during the fit results in a median value of $0.50 ^{+0.26}_{-0.09}$, which is in agreement with the core index range found within the FIRE simulation. For a fixed flux threshold of $\Phi(E >\SI{100}{\GeV}) = \SI{e-11}{\per \square \cm \per \s}$, we find about $N_\mathrm{BH}=11$ for $\gamma_\mathrm{c}=0.9$, $N_\mathrm{BH}=5$ for $\gamma_\mathrm{c}=0.3$, $N_\mathrm{BH}=7$ for $\gamma_\mathrm{c}$ as a free fitting parameter and $N_\mathrm{BH}=13$ if we assume the NFW profile. The number of IMBHs for a given flux threshold between our most optimistic scenario (NFW profile) and the most conservative scenario (cored profile with $\gamma_\mathrm{c}=0.3$) differs by a factor of $2.6$. Therefore, the choice of the dark matter density profile has a significant impact on the number of expected IMBHs surpassing a specific flux threshold. However, even in our most conservative scenario, gamma-ray observatories, such as H.E.S.S., provide a sensitivity that is sufficient to potentially detect $\sim 10$ IMBHs. In practise, the number of IMBHs that can be detected by ground-based gamma-ray observatories is limited by the number of IMBHs within the field of view of the experiment, as we discuss in more detail in Section~\ref{sec:discussion}.

\section{Dark matter annihilation channels}\label{sec:annihilation_channels}
\begin{figure}[htb]
    \centering
  
    \begin{subfigure}[b]{0.49\textwidth}
      \centering
      \includegraphics[width=\textwidth]{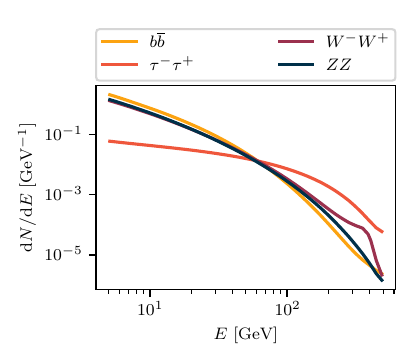}
    \end{subfigure}
    \hfill
    \begin{subfigure}[b]{0.49\textwidth}
      \centering
      \includegraphics[width=\textwidth]{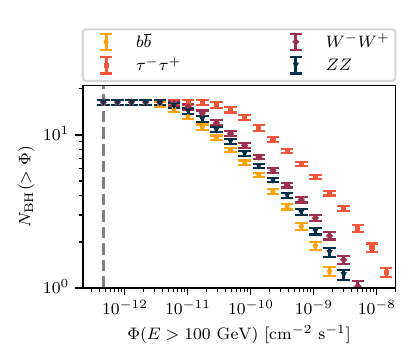}
    \end{subfigure}
  
    \caption{Left: Gamma-ray spectra per dark matter annihilation for different annihilation channels and $m_\chi = \SI{500}{\GeV}$. Right: Integrated luminosity of IMBHs for $\langle \sigma v \rangle = \SI{3e-26}{\cubic \cm \per \s}$ and $m_\chi = \SI{500}{\GeV}$ for different dark matter annihilation channels. The grey dashed line depictes the average H.E.S.S. flux sensitivity for the H.E.S.S. Galactic Plane survey~\cite{abdalla2018hess}.}
    \label{fig:annihilation_channels}
\end{figure}

In the main text we considered dark matter annihilation to occur via the $b\overline{b}$- and $\tau^-\tau^+$-channel. In the following, we briefly investigate the effect of different annihilation channels on the gamma-ray flux from dark matter self-annihilation around IMBHs. Figure~\ref{fig:annihilation_channels} (left) shows the gamma-ray spectra per dark matter annihilation for $m_\chi = \SI{500}{\GeV}$ and the $b\overline{b}$-, $\tau^-\tau^+$-, $W^-W^+$- and $ZZ$-channels. Generally, the $b\overline{b}$-, $W^-W^+$- and $ZZ$-channels result in fairly similar gamma-ray spectra with the most significant differences in the high energy regime ($\gtrsim \SI{100}{\GeV}$). All spectra decrease with increasing energy. The $\tau^-\tau^+$-channel leads to a lower gamma-ray emission compared to the other channels for $E \lesssim \SI{60}{\GeV}$ and to a significantly higher gamma-ray emission for $E \gtrsim \SI{60}{\GeV}$. Figure~\ref{fig:annihilation_channels} (right) shows the integrated luminosity of IMBHs for $E_\mathrm{th} = \SI{100}{\GeV}$, $\langle \sigma v \rangle = \SI{3e-26}{\cubic \cm \per \s}$ and $m_\chi = \SI{500}{\GeV}$ for the same four annihilation channels as in Figure~\ref{fig:annihilation_channels} (left). The $\tau^-\tau^+$-channel results in the highest gamma-ray fluxes, followed by the $W^-W^+$-, $ZZ$- and $b\overline{b}$-channel. This meets our expectation since we consider a energy threshold of $\SI{100}{\GeV}$ and the $\tau^-\tau^+$-channel leads to the highest number of gamma-ray photons per annihilation for $E \gtrsim \SI{60}{\GeV}$. Again, the  $b\overline{b}$-, $W^-W^+$- and $ZZ$-channels lead to fairly similar integrated luminosities as expected from the gamma-ray spectra. Considering a fixed flux threshold of $\Phi(E>\SI{100}{\GeV}) = \SI{e-10}{\per \square \cm \per \s}$, the $b\overline{b}$-channel results in $N_\mathrm{BH}=6$, the $ZZ$-channel in $N_\mathrm{BH}=7$, the $W^-W^+$-channel in $N_\mathrm{BH}=8$ and the $\tau^-\tau^+$-channel in $N_\mathrm{BH}=12$. Therefore, the $\tau^-\tau^+$-channel results in 2 times more IMBHs compared to the $b\overline{b}$-channel for a flux threshold of $\Phi(E>\SI{100}{\GeV}) = \SI{e-10}{\per \square \cm \per \s}$.

\newpage

\section{Galaxy catalogue of selected Milky Way-like galaxies}\label{app:galaxy}
\begin{table}[ht]
    \begin{center}
    \caption{Column descriptions for the galaxy catalogue of selected Milky Way-like galaxies}
    \label{tab:catalogue}
    \renewcommand{\arraystretch}{1.3}
    \begin{tabular}{lcp{10.5cm}}
        \hline
        Field & Unit & Description \\
        \hline
        \hline
        \code{galaxy\_id} & - & Unique identifier of the galaxy \\
        \code{group\_number} & - &  Integer identifier of the Friends-of-Friend (FoF) halo hosting this galaxy at $z=0$\\
        \code{subgroup\_number} & - &  Integer identifier of this galaxy within its FoF halo at $z=0$. The condition \code{subgroup\_number} = 0 selects central galaxies. \\
        \code{m} & $\mathrm{M}_\odot$ &  Total mass of the galaxy \\
        \code{m200} & $\mathrm{M}_\odot$ &  $M_{200}$ of the galaxy \\
        \code{m\_star} & $\mathrm{M}_\odot$ &  Stellar mas of the galaxy \\
        \code{m\_gas} & $\mathrm{M}_\odot$ &  Gas mass of the galaxy \\
        \code{sfr} & $\SI{}{\Msun \per \yr}$ & Star formation rate of the galaxy \\
        \code{fdisk} & - & Stellar disk-to-total mass ratio of the galaxy (values from~\cite{proctor2024identifying}) \\
        \code{fbulge} & - & Stellar bulge-to-total mass ratio of the galaxy (values from~\cite{proctor2024identifying}) \\
        \code{fihl} & - & Stellar IHL-to-total mass ratio of the galaxy (values from~\cite{proctor2024identifying}) \\
        \code{n\_sat} & - & Number of satellite galaxies associated with the galaxy \\
        \code{n\_sat\_stars} & - & Number of satellite galaxies with at least one star particle associated with the galaxy \\

    \end{tabular}
    \end{center}
\end{table}

\clearpage

\section{IMBH catalogue columns description}\label{app:imbh}
\begin{table}[ht]
    \begin{center}
    \caption{Column descriptions for the IMBH catalogue (part 1)}
    \label{tab:galaxy}
    \renewcommand{\arraystretch}{1.3}
    \begin{tabular}{lcp{10.5cm}}
        \hline
        Field & Unit & Description \\
        \hline
        \hline
        \code{main\_galaxy\_id} & - & Unique identifier of the main galaxy \\
        \code{host\_galaxy\_id} & - & Unique identifier of the host galaxy (differs only from \code{main\_galaxy\_id} if the IMBH is located in a satellite galaxy) \\
        \code{bh\_id} & - & Unique identifier of the black hole \\
        \code{m} & $\mathrm{M}_\odot$ & Black hole mass \\
        \code{z\_f} & - & Black hole formation redshift \\
        \code{z\_c} & - & Closest redshift for which a snapshot in the EAGLE simulations is available (see text for details) \\
        \code{nsnap\_c} & - & Closest snapshot for which a snapshot in the EAGLE simulations is available (see text for details) \\
        \code{d\_GC} & kpc & Distance of the black hole to the centre of potential of the host galaxy \\
        \code{lat\_GC} & rad & Galactic latitude of the black hole with the centre of potential of the host galaxy being the origin of the coordinate system \\
        \code{long\_GC} & rad & Galactic longitude of the black hole with the centre of potential of the host galaxy being the origin of the coordinate system \\
        \code{d\_Sun} & kpc & Distance of the black hole to the Sun \\
        \code{lat\_Sun} & rad & Galactic latitude of the black hole with the Sun being the origin of the coordinate system \\
        \code{long\_Sun} & rad & Galactic longitude of the black hole with the Sun being the origin of the coordinate system \\
        \code{m\_main\_galaxy} & $\mathrm{M}_\odot$ & Total mass of the main galaxy \\
        \code{m200\_main\_galaxy} & $\mathrm{M}_\odot$ & $M_{200}$ of the main galaxy \\
        \code{fdisk\_main\_galaxy} & - & Stellar disk-to-total mass ratio of the main galaxy (values from~\cite{proctor2024identifying}) \\
        \code{fbulge\_main\_galaxy} & - & Stellar bulge-to-total mass ratio of the main galaxy (values from~\cite{proctor2024identifying})  \\
        \code{fihl\_main\_galaxy} & - & Stellar IHL-to-total mass ratio of the main galaxy (values from~\cite{proctor2024identifying})  \\
        \code{m\_host\_galaxy} & $\mathrm{M}_\odot$ & Total mass of the host galaxy \\
        \code{m\_star\_host\_galaxy} & $\mathrm{M}_\odot$ & Stellar mass of the host galaxy \\
        \code{m\_gas\_host\_galaxy} & $\mathrm{M}_\odot$ & Gas mass of the host galaxy \\
        \code{sfr\_host\_galaxy} & $\SI{}{\Msun \per \yr}$ & Star formation rate of the host galaxy \\
    \end{tabular}
    \end{center}
\end{table}

\begin{table}[ht]
    \begin{center}
    \caption{Column descriptions for the IMBH catalogue (part 2)}
    \label{tab:imbh}
    \renewcommand{\arraystretch}{1.3}
    \begin{tabular}{lcp{10.5cm}}
        \hline
        Field & Unit & Description \\
        \hline
        \hline
        \code{gamma\_sp} & - & Spike index \\
        \code{r\_sp} & pc & Spike radius\\
        \code{rho(r\_sp)} & GeV/cm$^3$ & Dark matter density at the spike radius \\
         \code{satellite} & - & \code{True} if the black hole is located in one of the satellite galaxies of the host galaxy\\
         \code{n\_sat} & - & Number of satellite galaxies associated with the main galaxy of the IMBH \\
         \code{n\_sat\_stars} & - & Number of satellite galaxies with at least one star particle associated with the main galaxy of the IMBH \\
        \code{no\_host} & - & \code{True} if black hole was not assigned to any host at its formation redshift\\
        \code{r\_c} & kpc & Core radius (only available for cored dark matter profile) \\
    \end{tabular}
    \end{center}
\end{table}

\begin{table}[ht]
    \begin{center}
    \caption{Column descriptions for the flux catalogue, with each catalogue extracted for a designated dark matter mass.}
    \label{tab:fluxes}
    \renewcommand{\arraystretch}{1.3}
    \begin{tabular}{lcp{10.5cm}}
        \hline
        Field & Unit & Description \\
        \hline
        \hline
        \code{main\_galaxy\_id} & - & Unique identifier of the main galaxy \\
        \code{bh\_id} & - & Unique identifier of the black hole \\
        \code{sigma\_v} & $\SI{}{\cubic \cm \per \s}$ & Dark matter cross section times the relative velocity \\
        \code{r\_cut} & pc & Cutoff radius \\
        \code{flux} & $\SI{}{\per \square \cm \per \s}$ & Integrated gamma-ray flux for a given energy threshold $E_\mathrm{th}$, dark matter mass $m_\chi$, velocity weighted cross section $\langle \sigma v \rangle$ and annihilation channel \\
        
    \end{tabular}
    \end{center}
\end{table}

\end{document}